\documentclass[11pt,a4paper]{article}
\usepackage{jheppub}
\pdfoutput=1

\usepackage{amssymb,esvect,amsmath,graphicx,latexsym,amsthm,slashed,eso-pic,hyperref}

\usepackage{epsfig,amsmath,amssymb,verbatim,mathrsfs,hyperref}
\usepackage{xspace}
\usepackage{xcolor}
\usepackage{slashed}
\usepackage{dcolumn}
\usepackage{multirow}
\usepackage{color}
\graphicspath{{./}{./Figs/}{./Figures/}}

\def\beq{\begin{eqnarray}}
\def\eeq{\end{eqnarray}}
\def\bea{\begin{eqnarray}}
\def\eea{\end{eqnarray}}

\def\tev{\, {\rm TeV}}
\def\gev{\, {\rm GeV}}
\def\mev{\, {\rm MeV}}

\newcommand{\gsim}{\lower.7ex\hbox{$\;\stackrel{\textstyle>}{\sim}\;$}}
\newcommand{\lsim}{\lower.7ex\hbox{$\;\stackrel{\textstyle<}{\sim}\;$}}

\newcommand{\nnmb}{\nonumber}

\newcommand{\lrf}[2]{\left(\frac{#1}{#2}\right)}

\newcommand{\lag}{\mathscr{L}}

\newcommand{\kev}{\mathrm{keV}}

\newcommand{\zpr}{{0\prime}}
\newcommand{\mpr}{{-\prime}}
\newcommand{\onep}{{1_+}}
\newcommand{\onem}{{1_-}}

\begin{document}

\title{Relic Challenges for Vector-Like Fermions as Connectors to a Dark Sector}

\author[a,b,c]{Alexandre Carvunis,}
\author[c]{Navin McGinnis,}
\author[c]{and David E. Morrissey}

\affiliation[a]{Ecole normale supérieure Paris-Saclay, 4 Av. des Sciences, 91190 Gif-sur-Yvette, France}
\affiliation[b]{Laboratoire d’Annecy-le-Vieux de Physique Théorique, Université de Savoie Mont-Blanc et CNRS, B.P. 110, F-74941, Annecy Cedex, France}
\affiliation[c]{TRIUMF, 4004 Wesbrook Mall, Vancouver, BC, Canada V6T 2A3}

\emailAdd{a.carvunis@ip2i.in2p3.fr}
\emailAdd{nmcginnis@triumf.ca}
\emailAdd{dmorri@triumf.ca}

\abstract{
New dark sectors consisting of exotic fields that couple only
very feebly to the Standard Model~(SM) have strong theoretical
motivation and may be relevant to explaining the abundance 
of dark matter~(DM).  An important question for such sectors
is how they connect to the SM.  For a dark sector with a new 
gauge interaction, a natural connection arises from heavy vector-like
fermions charged under both the visible and dark gauge groups.
The gauge charges of such fermions imply that one or more of them is stable 
in the absence of additional sources of dark symmetry breaking. 
A generic challenge for such connectors is that they can produce 
too much dark matter or interact too strongly with nuclei if they 
were ever thermalized in the early universe.  
In this paper we study this challenge in a simple connector theory consisting 
of new vector-like electroweak doublet and singlet fermions that also transform
under the fundamental representation of a new (Abelian) gauge force,  
and we show that these connectors in their minimal form are almost always 
ruled out by existing direct DM searches.  To address this challenge,
we investigate two solutions. First, we study mitigating scattering
on nuclei by introducing a Majorana mass term for the singlet.
And second, we investigate a mixing with SM leptons that allows
the connectors to decay while remaining consistent with cosmological
tests and searches for charged lepton flavour violation. Both solutions
rely on the presence of a dark Higgs field with a specific charge.
}


\maketitle

\setcounter{page}{2}


\section{Introduction\label{sec:intro}}

  New gauge forces arise in many well-motivated extensions of 
the Standard Model~(SM)~\cite{Hewett:1988xc,Aldazabal:2000cn,Blumenhagen:2006ci,Langacker:2008yv}.  These forces are said to be dark if they do not 
couple directly to the matter of the SM~\cite{Alexander:2016aln}.  
Dark forces may also connect to dark matter~(DM) and play a crucial 
role in determining its properties and abundance 
today~\cite{Boehm:2003hm,Borodatchenkova:2005ct,Pospelov:2007mp,Arkani-Hamed:2008hhe,Pospelov:2008jd}.

Despite their name, dark forces are usually the most interesting when 
they are not completely dark and interact with 
the SM~\cite{Pospelov:2008zw,Bjorken:2009mm}. 
Abelian dark forces can connect to the SM through kinetic mixing 
with hypercharge~\cite{Okun:1982xi,Holdom:1985ag}, 
while non-Abelian dark forces typically require non-renormalizable 
connector operators~\cite{Faraggi:2000pv,Juknevich:2009ji,Juknevich:2009gg,Forestell:2017wov}. 
In both cases, a natural origin for the connector operators 
are new massive fermions that are charged under both the SM 
and dark gauge groups. For example, integrating out such states 
typically produces a contribution to the hypercharge kinetic mixing 
operator for an Abelian dark force~\cite{Okun:1982xi,Holdom:1985ag}, 
and dimension-eight $(F_{\mu\nu})^2(X_{\mu\nu})^2$ operators
in the both the Abelian and non-Abelian 
cases~\cite{Faraggi:2000pv,Juknevich:2009ji}. 
If the new massive matter states couple to the SM Higgs boson $H$, 
integrating them out can also connect the SM Higgs bilinear
to the dark vector field strength operator at dimension six,
$H^{\dagger}H\,(X_{\mu\nu})^2$~\cite{Juknevich:2009gg}.

While new connector fermions are well motivated, they can also be
problematic for cosmology if they are produced in the early universe. 
With only gauge interactions, such states generically have accidental 
flavour symmetries that make them stable. 
They can therefore lead to relic particles, possibly carrying charge 
under electromagnetism or the strong force.
Charged relics are clearly problematic~\cite{Perl:2009zz,DeLuca:2018mzn}, 
but even neutral relics with only weak charges can potentially 
obtain cosmological densities that are larger than the observed dark matter 
abundance or in conflict with bounds from direct and indirect detection. 
The objective of this work is to demonstrate the problem with a specific
minimal example and investigate some mechanisms to avoid it.

 The matter connectors that we consider consist of 
two exotic vector-like fermion multiplets with charge assignments under 
$SU(3)_c\times SU(2)_L\times U(1)_Y\times U(1)_x$
of $N = (1,1,0;\,q_x)$ and $P=(1,2,-1/2;\,q_x)$. 
This allows the Higgs Yukawa coupling and mass terms
\beq
-\lag \supset \left(\lambda\overline{P}\widetilde{H}N + h.c.\right)
+ m_P\overline{P}P + m_N\overline{N}N \ ,
\label{eq:vlep1}
\eeq
where $\widetilde{H} = i\sigma_2H^*$.  The new parameters $m_P$, $m_N$, 
and $\lambda$ can all be taken to be real and positive through field 
redefinitions.  We also normalize the dark gauge coupling $g_x$ 
such that $q_x= 1$.  
Coupling the connectors to the SM Higgs breaks what would otherwise
be independent flavor symmetries in each multiplet.
After electroweak symmetry breaking, 
the new fermions mix to form a pair of neutral Dirac fermions 
$\psi_1$ and $\psi_2$ as well as a charged Dirac fermion $P^-$.

 In the absence of an explicit $U(1)_x$ Higgs scalar, gauge invariance
implies that the lightest of the new connector states $\psi_1$
is stable; it will therefore contribute to the relic density 
of dark matter in the universethe universe.
Since the quantum numbers in this minimal theory force $\psi_1$ to be
a Dirac fermion, it can have a large spin-independent cross
section with nucleons from vector boson ($X$ and $Z$) exchange~\cite{Lee:1977ua}.
For fermion masses in the $100\,\gev$--$5\,\tev$ range, 
we show that this generically leads to unacceptably large rates 
in direct detection experiments, even if the $\psi_1$ relic density 
makes up only a small fraction of the total dark matter abundance. 

To address this challenge we investigate two extensions of this
minimal connector fermion model, discussed previously 
in Refs.~\cite{Davoudiasl:2012ig,Lu:2017uur} but
not studied in detail. In the first extension, we introduce
an explicit dark Higgs field $\Phi$ with $q_\Phi=-2q_x$ that develops a vacuum
expectation value and can induce a Majorana mass for the fermions.
This splits the neutral Dirac fermion states into pseudo-Dirac pairs with only
off-diagonal couplings to the vector bosons and eliminates the leading
contributions to nucleon scattering. We show that this can be sufficient
for consistency with current direct dark matter searches.
The second extension uses a  dark Higgs field $\phi$ with $q_\phi = q_x$
to mix the connector doublet with the lepton doublets of the SM. 
This coupling allows all the connector fermions to decay to SM states, 
but can also lead to charged lepton flavour violation~(LFV). We demonstrate that 
there exits a range of small couplings that are allowed by current LFV limits
and that also permit all the connector fermions to decay early enough
to avoid bounds from energy injection in the early 
universe~\cite{Kawasaki:2017bqm,Slatyer:2016qyl}.

 Our specific choice of connectors is motivated by obtaining a single 
potentially viable neutral relic particle, in contrast to electromagnetically
or strongly coupled relics. This choice can be generalized to larger
representations of $SU(2)_L$ with appropriate hypercharges~\cite{Cirelli:2005uq}.
Beyond breaking would-be flavor symmetries, the Higgs coupling is 
also motivated by other considerations. For example, this theory was
considered in relation to the muon magnetic moment anomaly 
in Ref.~\cite{Davoudiasl:2012ig} and as a mechanism for exotic Higgs decays 
in Refs.~\cite{Davoudiasl:2012ig,Lu:2017uur}, while similar Higgs couplings
to dark-charged vector-like fermions arise in many proposals to address 
the electroweak hierarchy 
problem~\cite{Azatov:2011ht,Heckman:2011bb,Graham:2015cka,Beauchesne:2017ukw}.
Such multiplets may also be expected in unification scenarios where 
the SM and $U(1)_x$ gauge groups are descendants of 
a single gauge group~\cite{Wojcik:2020wgm,Rizzo:2022lpm}.
We also note further related studies of connector fermions of 
Refs.~\cite{Cline:2016nab,Carone:2018eka,Rizzo:2018vlb,Kim:2019oyh,Lamprea:2019qet,Rueter:2019wdf,Coy:2020wxp}.

The outline of this paper is as follows. After this introduction we 
present in Sec.~\ref{sec:min} a minimal singlet-doublet theory 
of connector fermions and we study the laboratory bounds on the 
new states that it predicts. Next, in Sec.~\ref{sec:dmmin} we investigate
the thermal freezeout and dark matter signals of the stable $\psi_1$ fermion.
In Sec.~\ref{sec:maj} we present a simple extension of the theory with a Majorana
mass term that helps to alleviate the strong bounds we find
on the minimal theory from direct detection. 
We consider in Sec.~\ref{sec:lmix} a second extension of the minimal model 
that allows all the connector fermions to ultimately decay down
to the SM through lepton mixing and investigate the resulting implications.
Finally, Sec.~\ref{sec:conc} is reserved for our conclusions.
Some additional results are listed in an Appendix~\ref{sec:appa}.

\section{Review of the Minimal Theory and Laboratory Bounds
\label{sec:min}}

  We begin by studying the masses, interactions, and direct laboratory
bounds on the minimal fermionic connector theory consisting of 
vector-like $P=(1,2,-1/2;\,q_x)$ and $N=(1,1,0;\,q_x)$ fermions
charged under a new $U(1)_x$ gauge invariance with massive boson $X^\mu$. 
Our primary focus is on lighter dark vector bosons with $m_x \ll m_Z$.
We also normalize the dark gauge coupling such that $q_x=1$ without
loss of generality.

\subsection{Masses and Interactions}

 Electroweak symmetry breaking leads to mass mixing between
the neutral components of the $P$ and $N$ fermions.
Taking $H \to (0,v+h/\sqrt{2})$ in unitary gauge, 
the resulting spectrum of connector fermions consists of a charged fermion $P^-$ with mass $m_P$ from the doublet 
together with two neutral Dirac fermions $\psi_1$ and $\psi_2$ with masses
\beq
m_{1,2} =\frac{1}{2}\left[m_N+m_P\mp \sqrt{(m_N-m_P)^2+4\lambda^2v^2}\right] \ ,
\eeq
and mixing angles
\beq
\left(
\begin{array}{c}N\\P^0\end{array}
\right)
=
\left(
\begin{array}{cc}
c_\alpha&s_{\alpha}\\
-s_\alpha&c_\alpha
\end{array}
\right)
\left(
\begin{array}{c}\psi_1\\\psi_2\end{array}
\right) \ ,
\label{eq:psimix}
\eeq
where
\beq
\tan(2\alpha)= \frac{2\lambda v}{m_P-m_N} \ .
\eeq
We choose the solution for $\alpha$ such that $m_1 < m_2$.
The couplings of these states to the SM Higgs and electroweak
vector bosons are collected in Ref.~\cite{Lu:2017uur}.
Given the very strong direct bounds on light, 
electroweakly-charged fermions~\cite{Banta:2021dek},
we only consider parameters such that $m_1 > 0$ corresponding
to the condition $\sqrt{m_Nm_P} > \lambda\,v$.

Our theory also contains a new $U(1)_x$ vector boson $X^{\mu}$ with 
gauge coupling $g_x \equiv \sqrt{4\pi\,\alpha_x}$. 
We assume that this vector obtains a mass $m_x$ from either 
a dark Higgs~\cite{Pospelov:2007mp,Arkani-Hamed:2008hhe} 
or Stueckelburg mechanism~\cite{Stueckelberg:1957zz,Kors:2004dx}. 
In addition to its direct gauge
interactions with the new fermions (all with dark charge $q_x=1$ here),
the new vector connects with the SM through gauge kinetic mixing
with hypercharge~\cite{Okun:1982xi,Holdom:1985ag},
\beq
-\lag \ \supset \ \frac{\epsilon}{2\,c_W}\,B_{\mu\nu}X^{\mu\nu} \ .
\eeq
The mixing parameter $\epsilon$ allows the dark vetor to decay
and receives direct contributions from loops of the new $P$ fermions. 
For $m_1 \gg m_x$, which is the scenario we focus on here, 
the mediator fermions contribute an inhomogeneous term
to the renormalization group running of $\epsilon$ from a given ultraviolet
scale $\mu$ down to $m_P$ of
\beq
\Delta\epsilon
&\simeq& -\frac{1}{3\pi}\sqrt{\alpha\,\alpha_x}\,\ln\lrf{\mu}{m_P} \nnmb\\
&\simeq& -(3\times 10^{-3})\lrf{\alpha_x}{10\,\alpha}^{\!1/2}\ln\!\lrf{\mu}{m_P} \nnmb
\ .
\eeq
This suggests a natural size of the kinetic mixing on the order of 
$|\epsilon| \sim 10^{-3}$. Let us note, however, that the mixing can 
be eliminated or made parametrically small without fine tuning
by introducing a mirror copy of the connector fermions with opposite 
$U(1)_x$ charges and a small breaking of the mass degeneracy 
between them~\cite{DiFranzo:2015nli,DiFranzo:2016uzc,Gherghetta:2019coi}.
As a concrete benchmark for the dark vector in this work, 
we focus on $m_x \simeq 15\,\gev$ and $|\epsilon| \sim 10^{-4}$--$10^{-3}$
corresponding to a natural one-loop range.

  The massive fermions also generate an effective coupling of the SM Higgs 
boson to dark vectors at one-loop order. In the limit of $m_h \ll m_{1,2}$, 
the leading term is~\cite{Shifman:1979eb,Lu:2017uur}
\beq
\lag \ \supset \ \frac{\alpha_x}{6\pi}\,\frac{\lambda^2}{m_1m_2}\,H^{\dagger}H\,X_{\mu\nu}X^{\mu\nu} \ .
\label{eq:hopeff}
\eeq
A full expression for the loop function producing
this operator is given in Ref.~\cite{Lu:2017uur}.

\subsection{Laboratory Bounds on the Theory
\label{sec:bounds}}

\begin{figure}[ttt]
 \begin{center}
   \includegraphics[width = 0.47\textwidth]{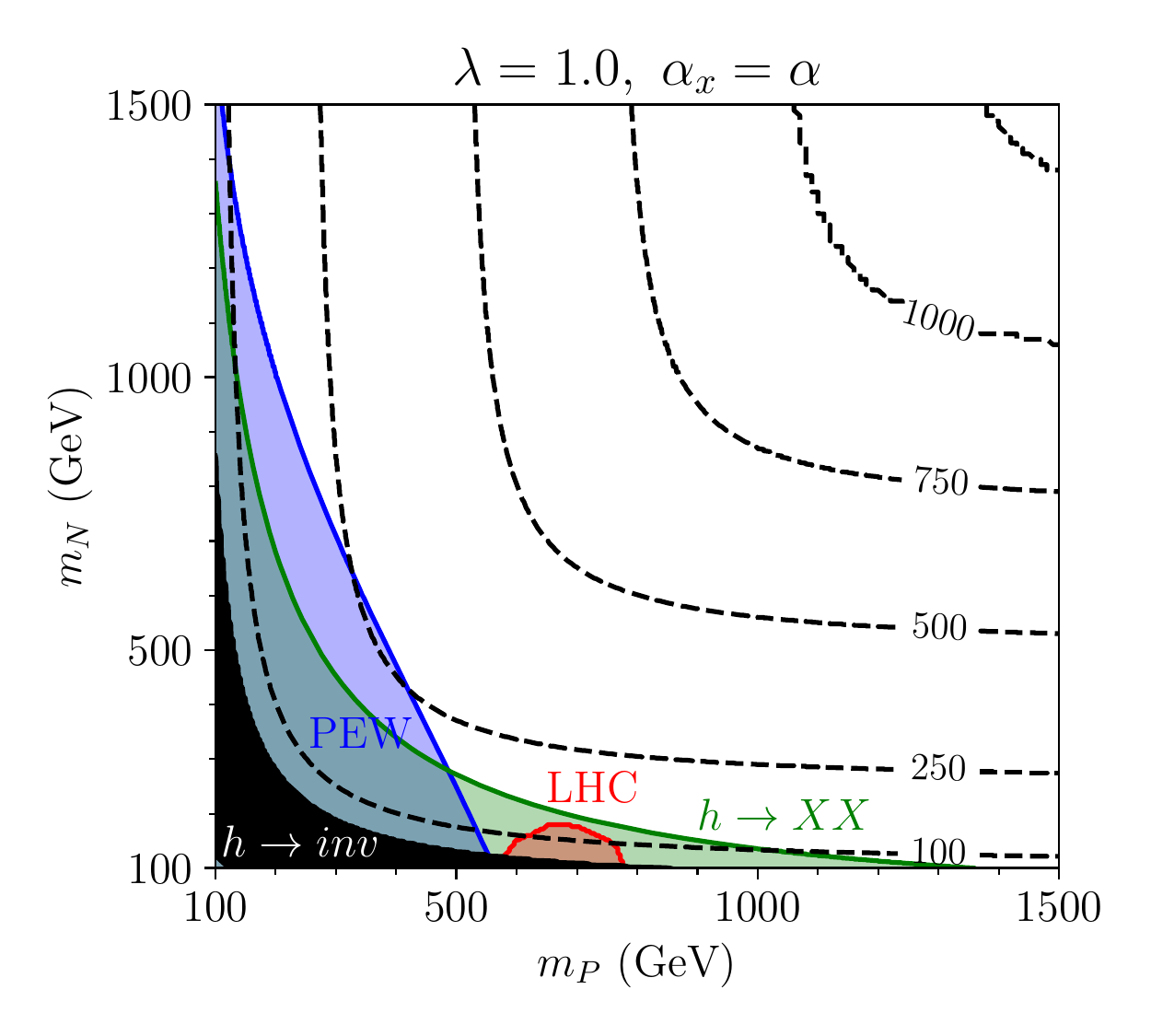}
   \includegraphics[width = 0.47\textwidth]{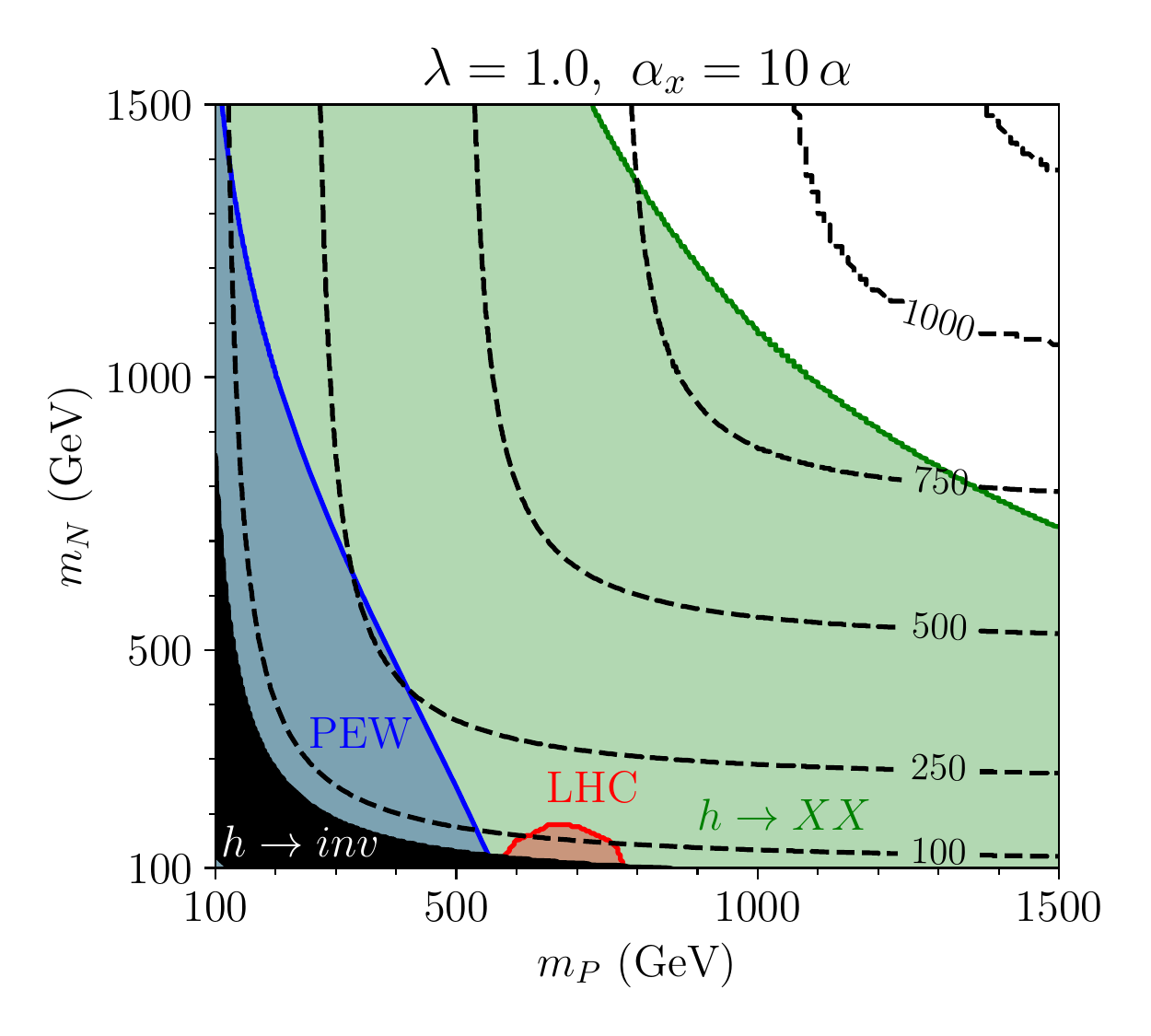}
 \end{center}
\vspace{-0.5cm}
 \caption{Laboratory bounds on the connector fermions in the $m_P$--$m_N$
plance for $\lambda=1.0$, and $\alpha_x=\alpha$~(left) 
and $\alpha_x=10\alpha$~(right). 
The solid black region is excluded by invisible Higgs boson decays~($h\to inv$),
the shaded green region indicates bounds from $h\to XX$ decays,
the blue region shows exclusions from precision electroweak tests~(PEW),
and the red region denotes bounds from direct searches at the LHC. 
The dashed lines are contours of constant $\psi_1$ masses $m_1$ in GeV.
}
 \label{fig:bounds}
 \end{figure}

 The new particles in the theory are constrained by direct tests of the 
dark vector boson, precision electroweak measurements,
collider searches at LEP and the LHC, and exotic Higgs decays.
We investigate the most important of these direct constraints here,
updating and expanding the earlier analysis of Ref.~\cite{Lu:2017uur}.
Our results are summarized in Fig.~\ref{fig:bounds}, which shows the exclusions
we find in the $m_P$--$m_N$ plane for $\lambda = 1.0$, 
together with $\alpha_x= \alpha$~(left) and $\alpha_x =10\,\alpha$~(right).
The dashed lines in this figure indicate contours of constant $\psi_1$ mass $m_1$.

For our benchmark dark vector mass of $m_x = 15\,\gev$, 
the strongest current direct bound on the dark vector~\cite{Agrawal:2021dbo} 
comes from the LHCb search for $X^\mu \to \mu^+\mu^-$ of Ref.~\cite{LHCb:2019vmc},
implying $|\epsilon| \lesssim 10^{-3}$ in this mass region. Searches
for dimuon resonances at CMS also provide a similar 
constraint~\cite{CMS:2019buh}. These limits are consistent with the range
of kinetic mixings we consider.

Mixing between the singlet and doublet can modify electroweak observables. 
The deviations induced are captured well by the oblique parameters
$S$, $T$, and $U$~\cite{Peskin:1990zt,Peskin:1991sw}. Full expressions
for the shifts in these parameters are given in Ref.~\cite{Lu:2017uur}.
We apply these results to compute $S$, $T$, and $U$ in the theory 
and compare them to the experimentally obtained values and correlations collected 
in Ref.~\cite{ParticleDataGroup:2020ssz} to derive exclusions.\footnote{
We use the Particle Data Group~\cite{ParticleDataGroup:2020ssz} 
combined value of the $W$ mass in our evaluation that does not include 
the recent, larger value obtained by the CDF 
collaboration~\cite{CDF:2022hxs}.}
These exclusions are shown in Fig.~\ref{fig:bounds} for $\lambda=1.0$. 

 The coupling of the $P$ and $N$ fermions to the Higgs field can give rise
to non-standard Higgs boson decay channels. If $m_1 < m_h/2$, 
the invisible decay $h\to \psi_1\bar{\psi}_1$ proceeds with partial width
\beq
\Gamma(h\to \psi_1\bar{\psi}_1) = 
\frac{\lambda^2\sin^2(2\alpha)}{16\pi}\,m_h\left[1-\lrf{2m_1}{m_h}^2\right]^{3/2} \ .
\eeq
Comparing to the SM Higgs width of $\Gamma_h \simeq 4.1\,\mev$,
the branching fraction of this channel easily exceeds the recent
ATLAS limit on invisible Higgs decays of 
$\text{BR}(h\to inv) < 0.145$~\cite{ATLAS:2022yvh} 
over essentially the entire parameter space we consider (with $\lambda\geq 0.1$).
The corresponding exclusions are shown in Fig.~\ref{fig:bounds}.

More visibly, loops of the heavy fermions give rise to 
$h\to XX$ decays which can produce highly distinctive pairs of
dilepton resonances. For $m_{1,2}\gg m_h$, this is described well by the
effective operator of Eq.~\eqref{eq:hopeff} yielding the partial width
\beq
\Gamma(h\to XX) = \frac{\alpha_x^2}{72\pi^3}\lrf{\lambda^2v^2}{m_1m_2}^{\!\!2}
\frac{m_h^3}{v^2}\left[1-\lrf{2m_x}{m_h}^{\!\!2}+6\lrf{m_x}{m_h}^{\!\!4}\right]\,
\sqrt{1-\lrf{2m_x}{m_h}^{\!\!2}}.~~
\eeq
This decay channel has been searched for by ATLAS~\cite{ATLAS:2021ldb} 
and CMS~\cite{CMS:2021pcy},
with the former analysis giving the most stringent bound for $m_x = 15\,\gev$
of $\text{BR}(h\to XX) < 2.35\times 10^{-5}$. The decay can also be mediated
directly by kinetic mixing~\cite{Curtin:2014cca}, but the associated branching 
ratio scales like $\epsilon^4$ and is strongly subleading relative to 
the loop decay for the parameters considered here. 
Bounds from $h\to XX$ are given in Fig.~\ref{fig:bounds}, and are particularly
strong for larger $\lambda$ and $\alpha_x$, but fall off quickly for 
smaller $\lambda$ or $\alpha_x$ (scaling like $\alpha_x^2\lambda^4$). 
Loops as well as kinetic mixing also give rise to $h\to XZ$, 
but we find that the resulting constraints are
not as strong as those from $h\to XX$. 

Collider searches are sensitive to the heavier connector fermions
to the extent that they can be created efficiently. 
In the minimal doublet-singlet $P$--$N$ model considered in this work, 
the production and decay channels are analogous to a Higgsino-Bino 
system in the minimal supersymmetric Standard Model~\cite{Canepa:2020ntc} 
and the associated collider signals are very similar~\cite{Cohen:2011ec,Martin:2014qra,Liu:2020muv}. 
Based on this analogy, collider bounds on the theory from LEP and the LHC 
were estimated in Ref.~\cite{Lu:2017uur}. 
Results from LEP constrain $m_P\gtrsim 90$--$100\,\gev$ depending 
on the mass splitting between the $P^\pm$ and $\psi_1$ states. 
Here, we focus on $m_P,\,m_N \geq 100\,\gev$ and derive updated limits 
from new LHC data.

Pair production of the connector fermions at the LHC occurs predominantly 
through $s$-channel Drell-Yan processes with $\gamma /Z$ or $W$ boson exchanges. 
The rates for these processes therefore depend on the doublet 
content of the relevant states. Once created, the connector fermions decay 
down to the lightest state $\psi_1$ through $P^{\mp}\to W^{\mp\,(*)}\psi_1$ 
or $\psi_2 \to h/Z^{(*)}\psi_1$. These considerations imply that up 
to mixing effects, the fermions are the most detectable for the mass 
hierarchy $m_N < m_P$. The $P^{\mp}$ and $\psi_2$ states are then analogous
to the $\chi_1^{\mp}$ and $\chi_{2,3}^0$ states of a Higgsino-Bino system
for $|M_1| < |\mu|$, with LHC signatures involving significant missing
energy and the decay products of the electroweak bosons. When $m_P < m_N$,
the mostly-doublet states tend to be close in mass and their decay products
are soft and difficult to detect.

 Recently the ATLAS collaboration has performed a comprehensive search 
for the chargino-neutralino system that targets Higgsino-Bino signals~\cite{ATLAS:2021yqv}. 
To estimate the bounds on connector fermions implied by this search we calculate 
production cross sections using {\tt{MadGraph5}}~\cite{Alwall:2014hca} 
where the couplings and mass eigenstates were implemented using 
{\tt{FeynRules\;2.3}}~\cite{Alloul:2013bka}. We consider all possible 
decay chains of $P^{\pm}$ and $\psi_2$ into EW and Higgs bosons 
and estimate the corresponding strength of each signal in the signal regions 
identified as 4Q-VV, 2B2Q-WZ, and 2B2Q-Wh in the analysis of 
Ref.~\cite{ATLAS:2021yqv} using the datasets of efficiencies 
and detector acceptances available from the HEPData page 
of the corresponding search~\cite{hepdata.104458}. 
Using the provided SM backgrounds for each region, the combined significance 
is then calculated for a given $\lambda, m_{P},$ and $m_{N}$ using the asymptotic
approximations given in~\cite{Cowan:2010js}. We show the resulting 
95\% CL exclusion regions in the $m_{P}$-$m_{N}$ plane for $\lambda=1.0$
in Fig.~\ref{fig:bounds}. Our results align closely with those presented 
in~\cite{ATLAS:2021yqv} for the Higgsino-Bino system as expected.

  The collected model bounds shown in Fig.~\ref{fig:bounds} correspond 
to the relatively large value $\lambda = 1.0$. As $\lambda$ decreases,
many of these bounds weaken significantly. For $\lambda = 0.1$,
we find that the only remaining bound in the parameter space shown comes
from direct LHC searches and is nearly identical to the $\lambda=1.0$ case. 
Let us also point out that large values of $\lambda$ 
tend to destabilize the Higgs potential by driving the Higgs quartic coupling
negative at a lower scale than in the SM. This effect was studied in 
Ref.~\cite{Lu:2017uur}, and for the parameters considered in this work the scenario can be treated as a self-consistent effective theory up to
energies of at least $5\,\tev$.

\section{Dark Matter in the Minimal Connector Theory~\label{sec:dmmin}}

The analysis in the previous section shows that the new connector fermions 
are safe from direct collider searches and precision electroweak tests 
for masses $m_1 \gtrsim 100$--$700,\gev$, depending on the specific 
mass spectrum and coupling strengths. 
Even stronger bounds can be derived on the theory from the contribution
of the lightest stable new fermion $\psi_1$ to the abundance of dark matter 
and signals in direct dark matter searches. In particular, the vector couplings
of this fermion lead to direct detection rates that are nearly always ruled
out, even when this state only makes up a small fraction of the total 
dark matter density.

\subsection{Relic Densities}

\begin{figure}[ttt]
 \begin{center}
   \includegraphics[width = 0.47\textwidth]{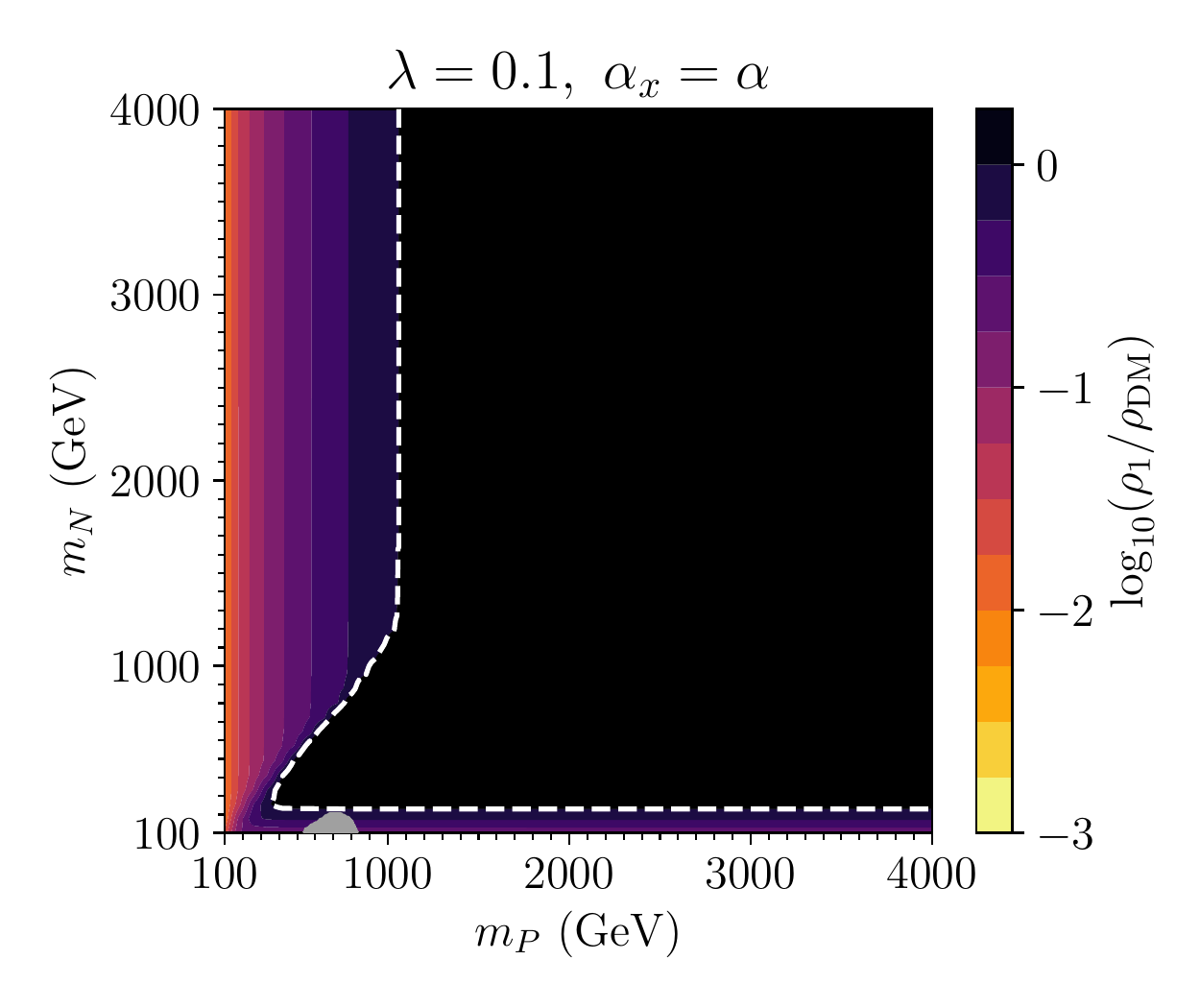}
   \includegraphics[width = 0.47\textwidth]{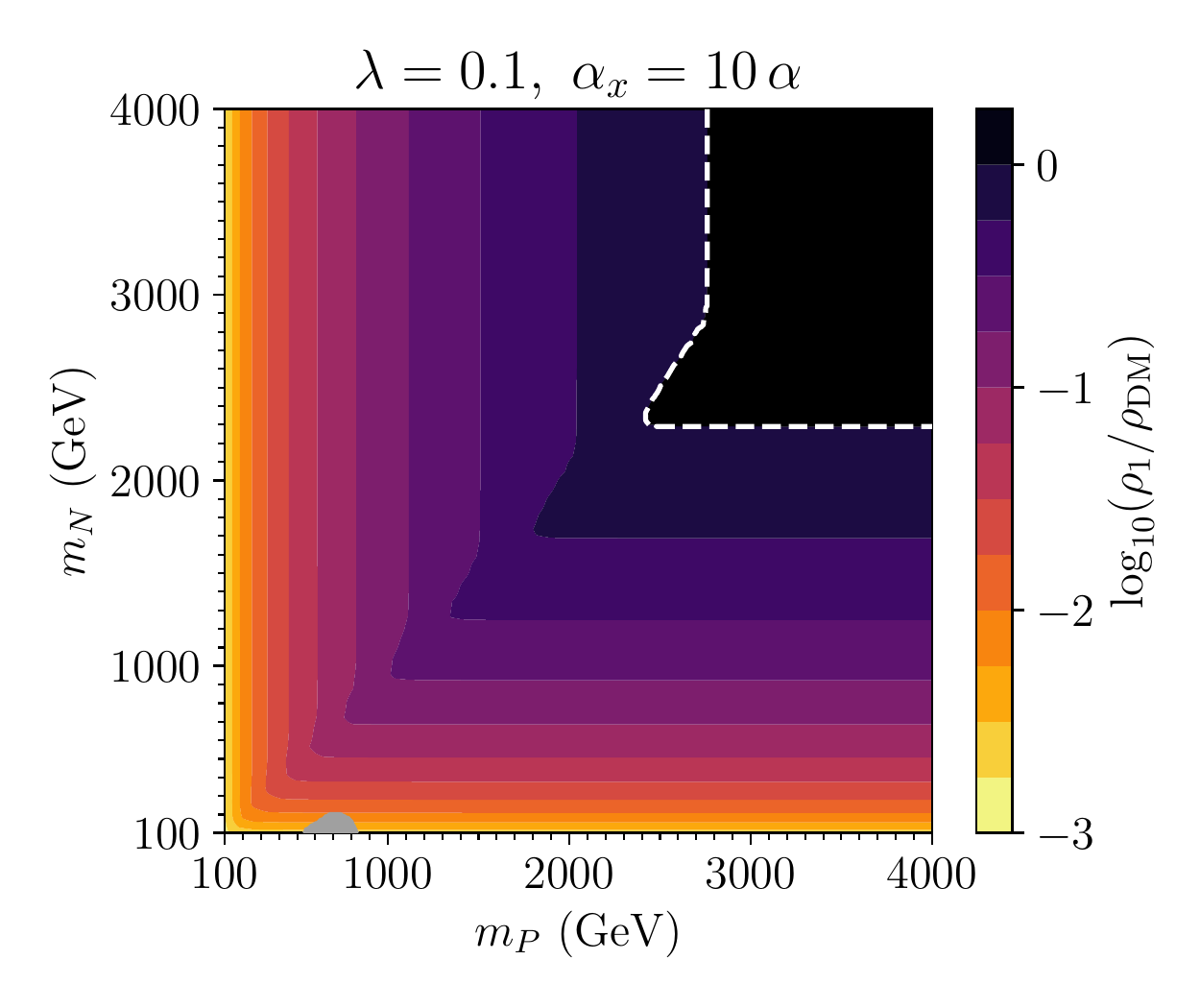}\\
   \includegraphics[width = 0.47\textwidth]{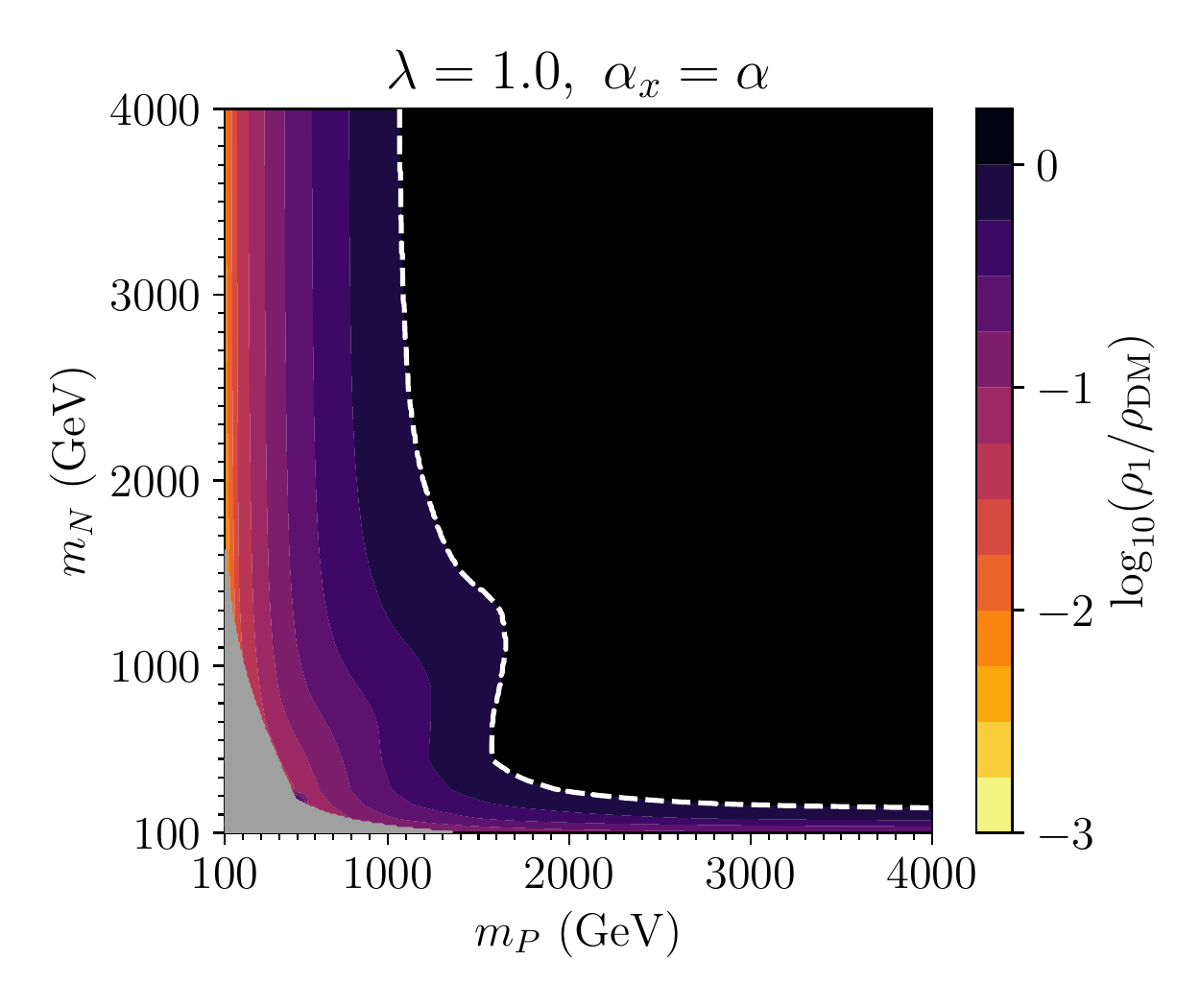}
   \includegraphics[width = 0.47\textwidth]{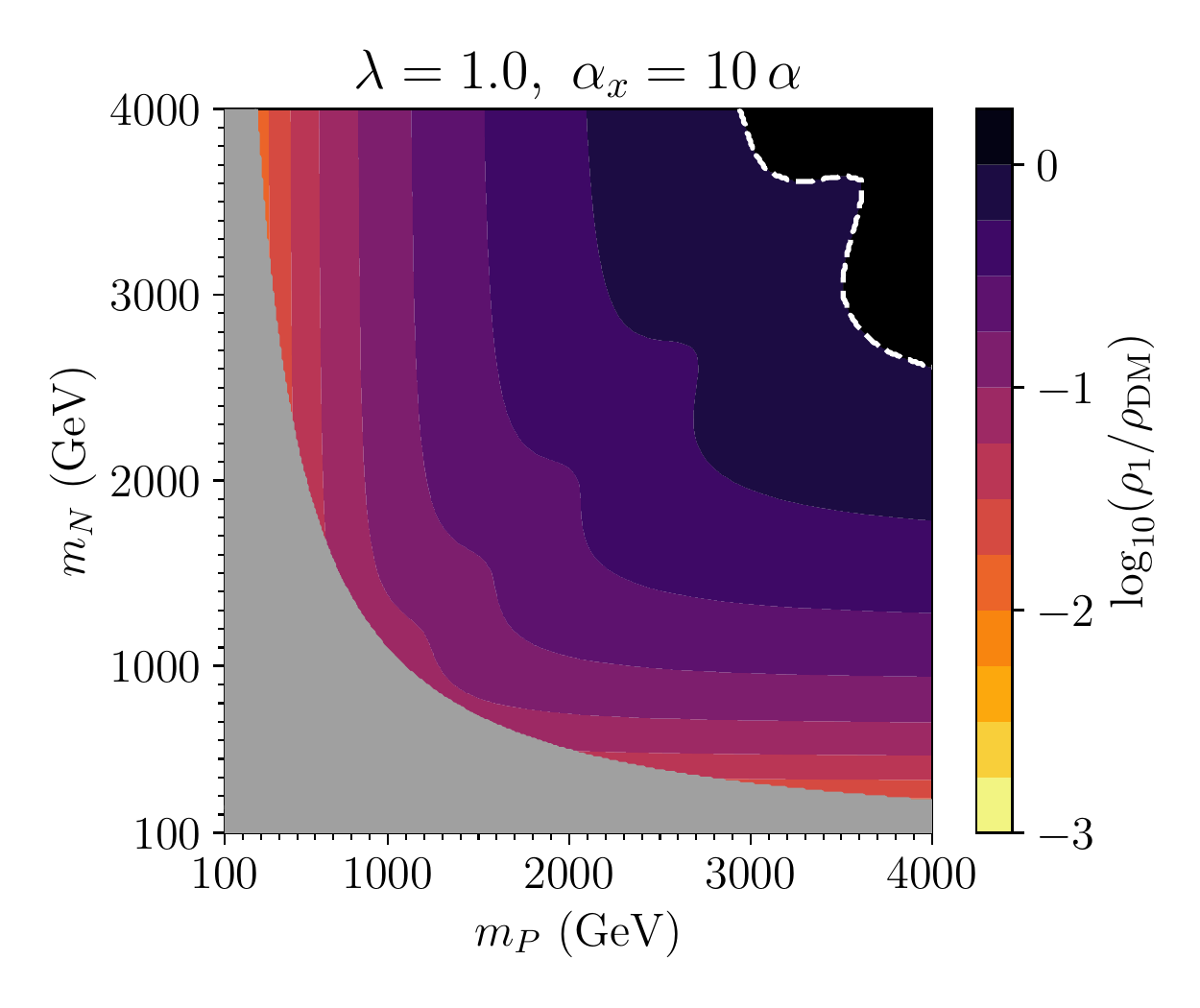}
 \end{center}
\vspace{-0.5cm}
 \caption{
Fractional relic densities $\rho_1/\rho_\text{DM}$ 
of the $\psi_1$ connector fermion for
$\lambda = 0.1$~(top) and $\lambda=1.0$~(bottom) 
with $\alpha_x = \alpha$~(left) and $\alpha_x= 10\,\alpha$~(right).
The grey shaded regions show the combined exclusions from direct searches
while the black shaded regions indicate where $\psi_1$
produces too much thermal dark matter.
}
 \label{fig:reld}
 \end{figure}

Thermal reactions in the early universe would have created the exotic fermions
$\psi_1$, $\psi_2$, and $P^\mp$ with significant abundances if the 
temperature was ever hot enough, $T \gtrsim m_1/20$. The fermions would have
then undergone thermal freezeout as the universe cooled to produce a relic 
density of neutral $\psi_1$ particles. We assume that this occured with 
no subsequent large injections of entropy to dilute their relic density
beyond the standard expansion of spacetime.

 To compute the $\psi_1$ relic abundance, we have implemented the theory
in {\tt{FeynRules~2.3}}~\cite{Christensen:2008py,Degrande:2011ua,Alloul:2013bka} 
and interfaced it with {\tt{MadDM~v3.2}}~\cite{Backovic:2013dpa,Backovic:2015cra,Ambrogi:2018jqj}. 
We show the resulting $\psi_1$ relic densities $\rho_1$
relative to the observed dark matter abundance $\rho_\text{DM}$ 
in Fig.~\ref{fig:reld} as a function of the mass parameters $m_P$ and $m_N$ 
for $\lambda = 0.1$~(top) and $\lambda =1.0$~(bottom), 
together with $\alpha_x=\alpha$~(left) and $\alpha_x=10\alpha$~(right). 
For these coupling values, we expect that non-perturbative enhancements 
of the annihilation cross section at
freeze-out will be negligible to mild~\cite{Arkani-Hamed:2008hhe,Cassel:2009wt,Slatyer:2009vg}.
The grey shaded regions in these plots summarize
the exclusions from direct laboratory searches as discussed 
in Sec.~\ref{sec:bounds}. 

Since we focus on the limit $m_1 \gg m_x$, 
while direct bounds typically require $m_1\gtrsim m_Z$, 
the $\psi_1$ state always has efficient annihilations to at least some
vector bosons in the parameter regions of interest, $\psi_1+\bar{\psi}_1 \to V+V$ 
where $V=X,Z,W$. When $m_P < m_N$ and
$m_P \gg \lambda\,v$, the $\psi_1$ and $P^-$ masses tend to be close
to each other and coannihilation can be significant.
These features are evident in Fig.~\ref{fig:reld}.
For larger $\alpha_x=10\,\alpha$, DM annihilation is dominated
by $\psi_1+\bar{\psi}_1\to X+X$ and the relic abundance depends
mainly on $m_1$. In contrast, for smaller $\alpha_x=\alpha$
annihilations to weak vector bosons become important and we see
a more efficient depletion of the relic density when $\psi_1$
is composed mainly of $P^0$ ($m_P< m_N$) relative to when it
is mostly $N^0$~($m_N < m_P$). Indeed, with $m_P\gg m_N$ the $\psi_1$
state is analogous to a pure Higgsino lightest superpartner in supersymmetry
with $\mu \sim m_P \sim m_1$, and the correct relic density is obtained
for the familiar value of $m_P\simeq 1100\,\gev$~\cite{Mizuta:1992qp,Cirelli:2007xd,Dessert:2022evk}.

The relic densities shown in Fig.~\ref{fig:reld} generally grow larger
as the mass $m_1$ increases. Even for larger dark-sector couplings
$\alpha_x=10\,\alpha$, the $\psi_1$ relic produces too much dark matter
if $m_1 \gtrsim 3\,\tev$. This illustrates the generic challenge of
such connector fermions and provides a further motivation for them
to not be too heavy, beyond considerations of naturalness.

\subsection{Constraints from Direct Detection}

\begin{figure}[ttt]
 \begin{center}
   \includegraphics[width = 0.47\textwidth]{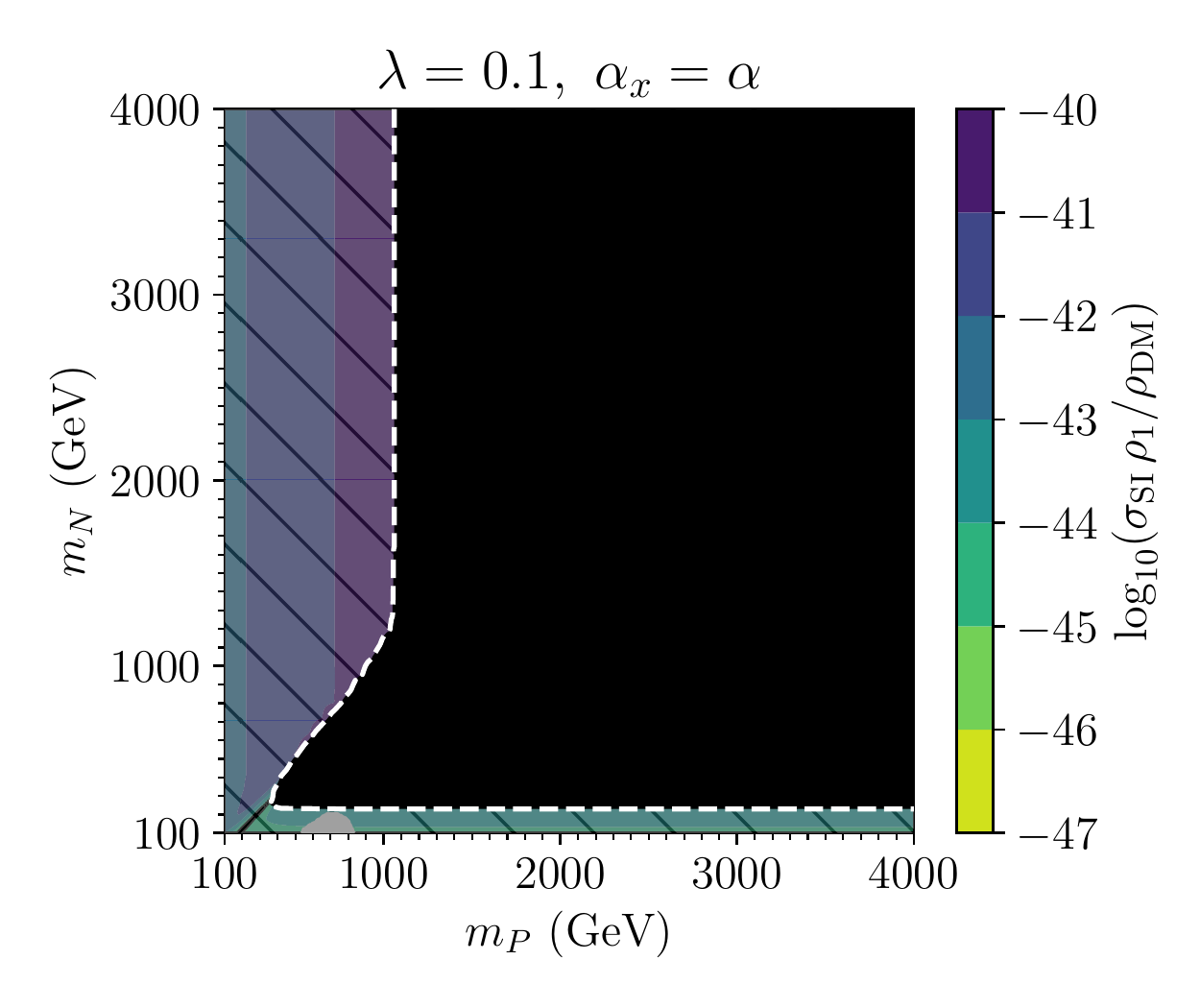}
   \includegraphics[width = 0.47\textwidth]{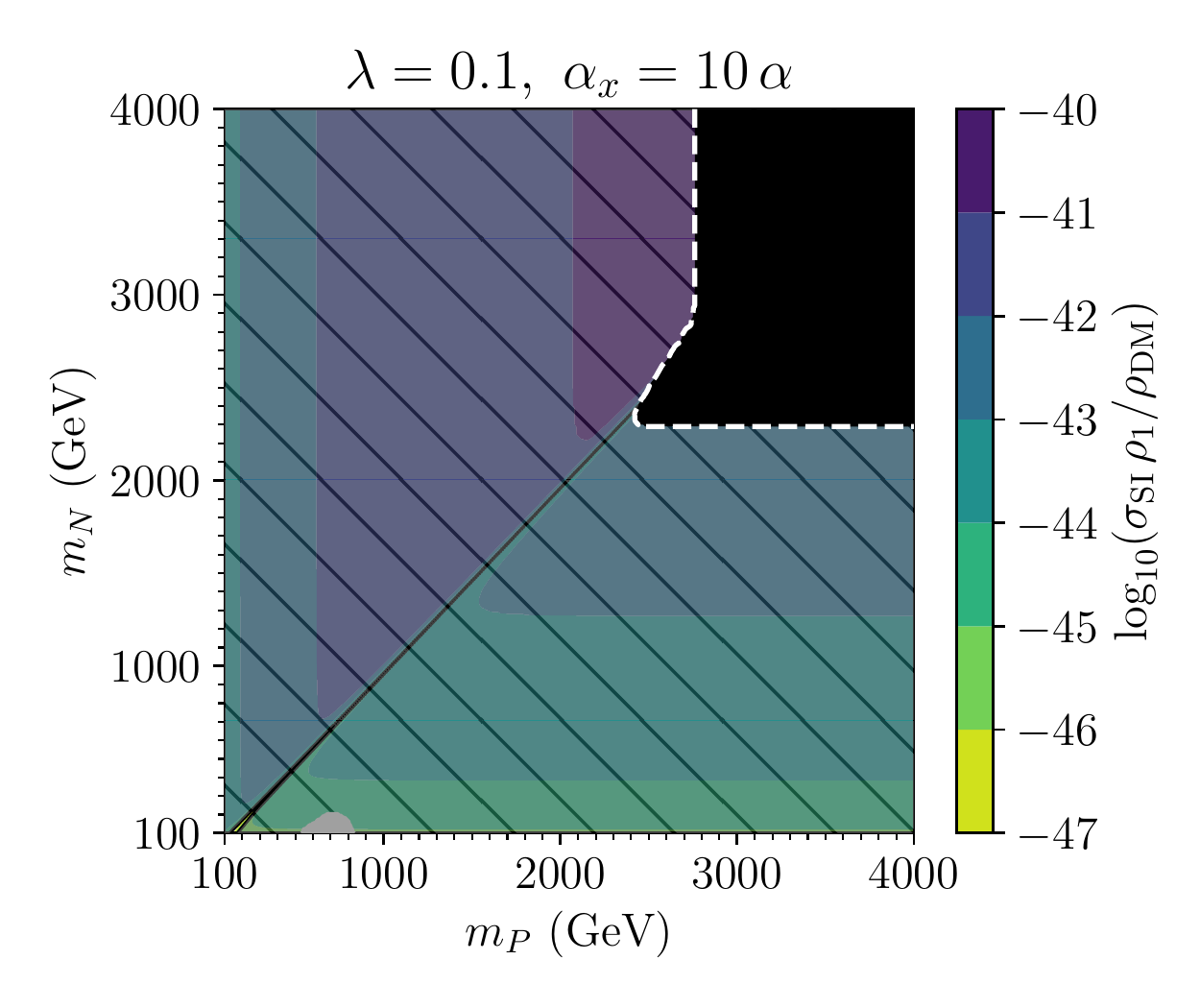}\\
   \includegraphics[width = 0.47\textwidth]{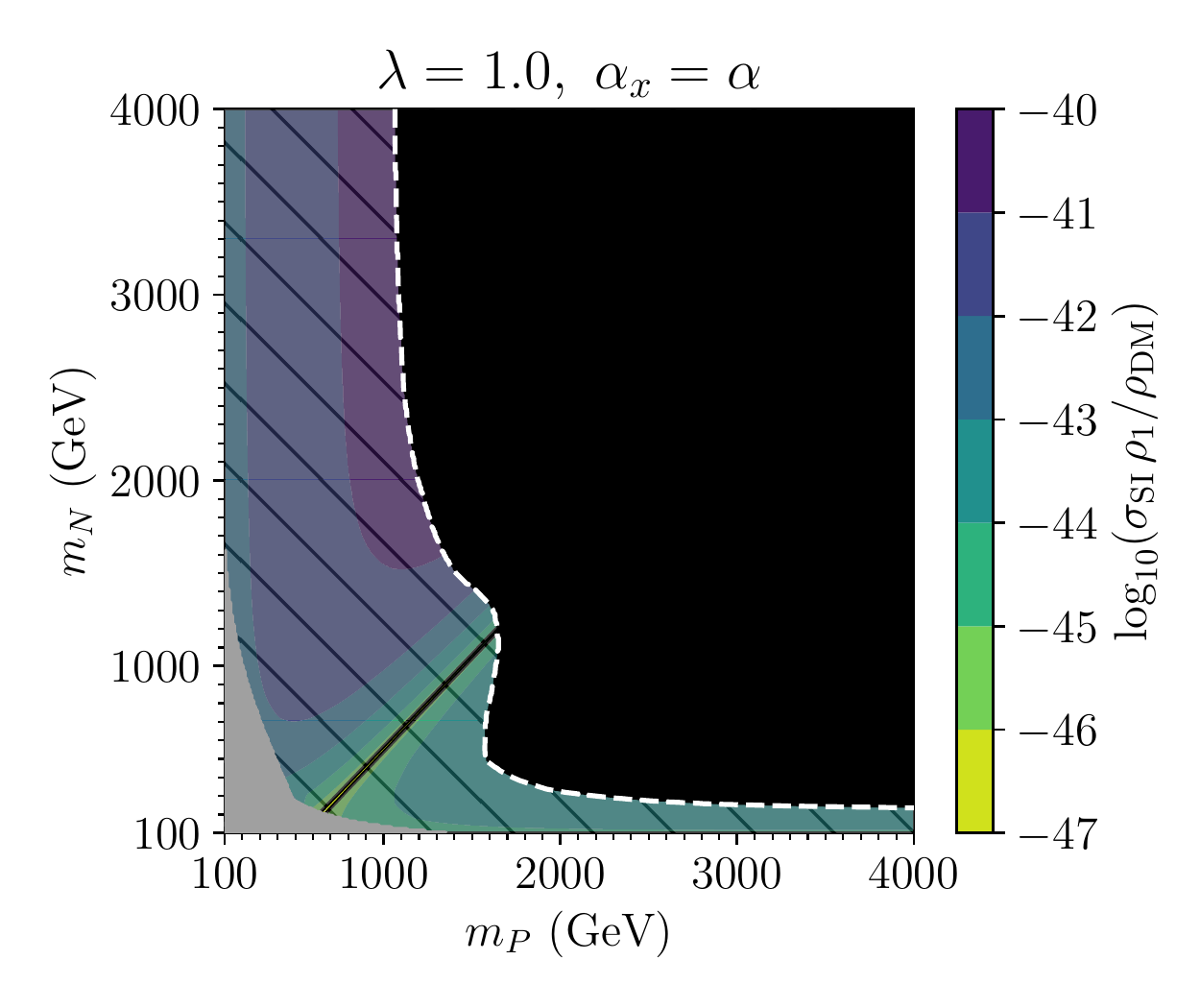}
   \includegraphics[width = 0.47\textwidth]{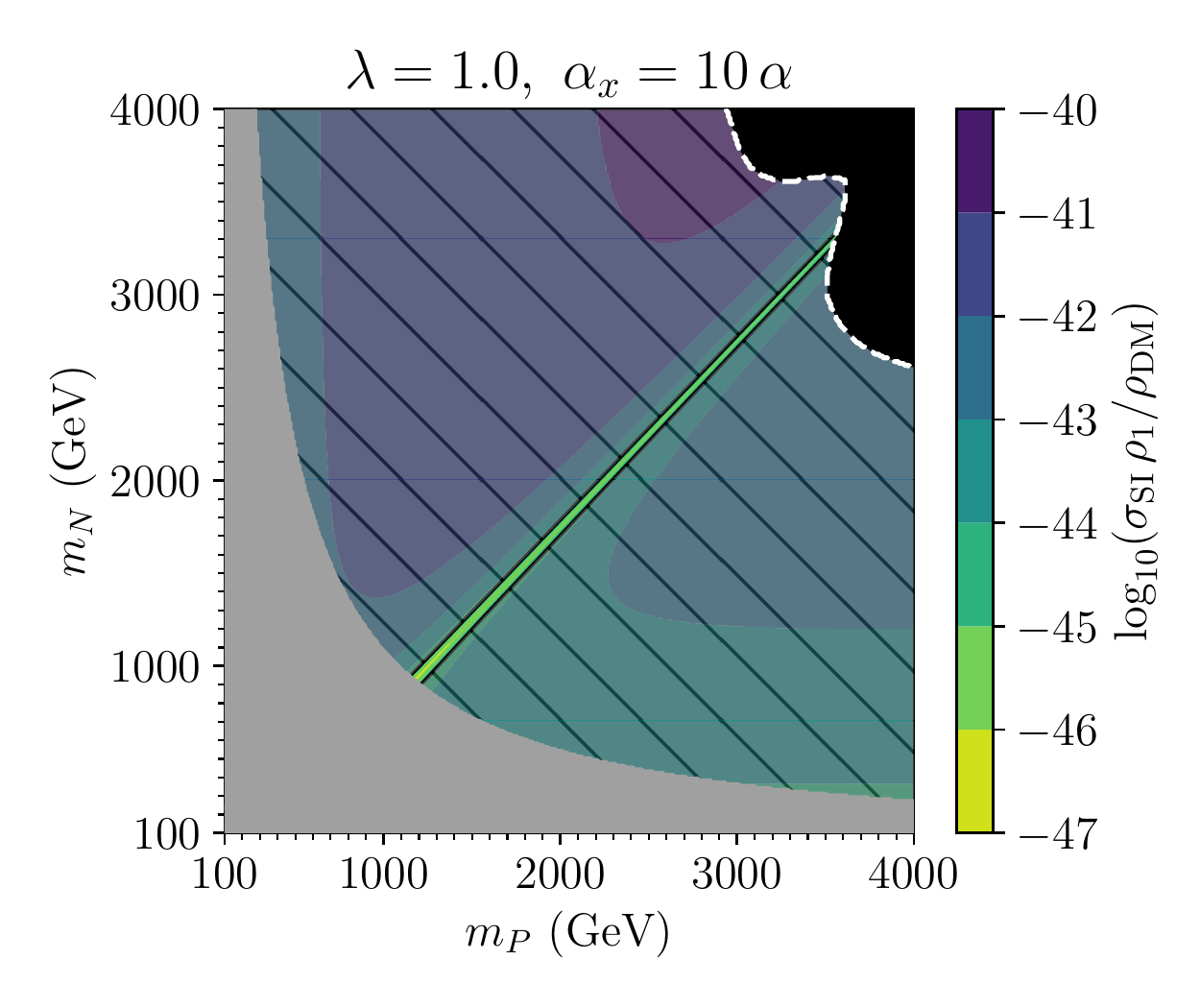}
 \end{center}
\vspace{-0.5cm}
 \caption{Contours of the density-weighted spin-independent
per nucleon cross section $(\sigma_\text{SI}\,\rho_1/\rho_\text{DM})$ 
for the minimal connector theory
in the $m_P$--$m_N$ plane for $\lambda = 0.1,\,1.0$~(top and bottom) 
and $\alpha_x = \alpha,\,10\,\alpha$~(left and right).
The downward hatched regions are excluded by DM direct detection
searches, the grey shaded regions show the combined exclusions 
from direct searches, and the black regions indicate where $\psi_1$
produces too much thermal dark matter.}
 \label{fig:dd1}
 \end{figure}

 While the lightest connector fermion $\psi_1$ can have an acceptably
small relic density, it is a Dirac fermion that interacts
with nuclei with an unsuppressed spin-independent interaction mediated by the
$Z$ and $X$ vector bosons. This can lead to very large cross sections on
nuclei that are in tension with searches from direct detection experiments,
even when the $\psi_1$ states makes up only a very small fraction of the 
total DM abundance~\cite{Davoudiasl:2012ig,Lu:2017uur}.

The leading effective per-nucleon spin-independent~(SI) cross section
on a target nucleus $N=(A,Z)$ is~\cite{Jungman:1995df}
\beq
\sigma_\text{SI} = \frac{\mu_n^2}{\pi}\left[\frac{Zf_p+(A-Z)f_n}{A}\right]^2 \ ,
\label{eq:sigsi}
\eeq
where $\mu_n = m_nm_1/(m_n+m_1)$ is the DM-nucleon reduced mass, and
\beq
f_p &=& 
\frac{G_F}{\sqrt{2}}s_{\alpha}^2(1-4s_W^2) 
- \frac{4\pi}{m_x^2}\epsilon\,\sqrt{\alpha\alpha_x}
-\tilde{d}_p\left[\frac{2}{9}+\frac{7}{9}\sum_qf_{q}^{p}\right] \ ,\\
f_n &=& -\frac{G_F}{\sqrt{2}}s_{\alpha}^2+0
-\tilde{d}_n\left[\frac{2}{9}+\frac{7}{9}\sum_qf_{q}^{n}\right] \ .
\eeq 
The three terms in each of these quantities correspond to $Z$, $X$, and Higgs
exchange, respectively. For the Higgs terms, the sums run over light quarks 
$q=u,d,s$, the coefficients $f_{q}^{p,n}$ are collected in Refs.~\cite{DelNobile:2021wmp},
and the couplings $\tilde{d}_{p,n}$ are given by
\beq
\tilde{d}_{p} = \tilde{d}_n = -\frac{m_{n}}{2\,v}\frac{\lambda\,\sin(2\alpha)}{m_h^2}
\ ,
\eeq
for the Higgs mass $m_h$. 

 Combining these expressions with the relic densities above, we can estimate
the experimental bounds on the connector fermions from direct detection
experiments. For spin-independent scattering in the mass range of interest, 
$m_1 \gtrsim 100\,\gev$,
the most stringent current limits come from 
LUX-ZEPLIN~\cite{LUX-ZEPLIN:2022qhg}. 
For a given $\psi_1$ mass, we compare the stated limit on the 
per-nucleon cross section $\sigma_\text{SI}$ to the density-weighted
value $(\rho_1/\rho_\text{DM})\,\sigma_\text{SI}$ derived here assuming 
a xenon target.

In Fig.~\ref{fig:dd1} we show contours of the density-weighted, spin-independent
per nucleon cross section $\sigma_\text{SI}\,\rho_1/\rho_\text{DM}$ 
in the $m_P$--$m_N$ plane for $\lambda = 0.1,\,1.0$~(top and bottom) 
and $\alpha_x = \alpha,\,10\,\alpha$~(left and right). 
We also fix the kinetic mixing at $\epsilon = -10^{-4}$ for reference. 
The downward hatched regions in these panels show the current exclusions
from direct detection searches~\cite{LUX-ZEPLIN:2022qhg}.
As before, the filled grey regions show the combined bounds from laboratory
searches for the connector fermions while the black regions
indicate where the thermal relic density of $\psi_1$ exceeds
the observed value.

  Nearly the entire parameter region consistent with laboratory searches
and the $\psi_1$ relic density shown in the panels of Fig.~\ref{fig:dd1}
is ruled out by direct detection searches. This is despite the $\psi_1$
relic density often only making up a very small fraction of the total
dark matter density. The strong exclusions come from the direct vector
couplings of $\psi_1$ to nucleons through the $Z^0$ and $X$ vector bosons,
which avoid the suppression by small Yukawa couplings or loops seen
in Higgs boson exchange. The only parameter region not excluded is a very
thin sliver where the various contributions the effective SI cross section
cancel almost completely. This occurs for negative $\epsilon$,
which we have chosen here to illustrate the effect. We conclude from
Fig.~\ref{fig:dd1} that the minimal connector fermion scenario is
all but ruled out assuming a standard, thermal cosmological history.

\section{Moderated Signals from a Majorana Mass Term\label{sec:maj}}  

Having found that the connector fermions in the theory are nearly
completely excluded by direct dark matter searches, we turn now to
ways to mitigate their impact on direct detection.
As a first approach, we expand the minimal theory with an explict
dark Higgs field $\Phi$ with $U(1)_x$ charge $q_\Phi = -2q_x$. 
This allows the new Yukawa coupling
\beq
-\lag \ \supset \ \frac{1}{2}y_N\,\Phi\,\overline{N^c}N + h.c.
\label{eq:mmaj}
\eeq
where we have given equal masses to both Weyl components for simplicity.
If $\Phi$ develops a VEV, $\langle \Phi\rangle = \eta$,
the new coupling generates a Majorana mass term $M = y_N\eta$ 
for the SM-singlet $N$ fermion.
These split the two Weyl components of the Dirac fermion
$\psi_1$ into a pair of Majorana fermions with mass splitting $\Delta m$.
Such a splitting implies exclusively off-diagonal couplings to the $Z$ and $X$
vector bosons~\cite{Tucker-Smith:2001myb}, 
which strongly suppresses nuclear scattering
through vector exchange for 
$\Delta m \gtrsim 200\,\kev$~\cite{Bramante:2016rdh,PandaX:2022djq}.
Instead, the dominant contribution to nuclear scattering comes
from Higgs exchange via the mixing coupling $\lambda$.

\subsection{Majorana Mass Splittings}

Starting from the mass eigenstate basis in the minimal theory
with Dirac fermions $\psi_1$ and $\psi_2$, it is convenient to re-express
them in terms of their Weyl components as
\beq
\psi_1 = \left(\begin{array}{c}\chi_1\\\bar{\chi}_1^c\end{array}\right) \ ,
\qquad 
\psi_2 = \left(\begin{array}{c}\chi_2\\\bar{\chi}_2^c\end{array}\right) \ .
\eeq
The mass matrix in the basis $(\chi_1,\chi_1^c,\chi_2,\chi_2^c)^t$ is then
\beq
\mathcal{M} = \left(
\begin{array}{cccc}
c_\alpha^2\,M&m_1&s_\alpha c_\alpha M&0\\
m_1&c_\alpha^2\,M&0&s_\alpha c_\alpha\,M\\
s_\alpha c_\alpha M&0&s_\alpha^2\,M&m_2\\
0&s_\alpha c_\alpha\,M&m_2&s_\alpha^2\,M
\end{array}
\right) \ .
\eeq
This matrix is diagonalized by the orthogonal transformation $\mathcal{O}$
given by
\beq
\mathcal{O} = 
\left(\begin{array}{cccc}
\frac{1}{\sqrt{2}}&\frac{1}{\sqrt{2}}&0&0\\
-\frac{1}{\sqrt{2}}&\frac{1}{\sqrt{2}}&0&0\\
0&0&\frac{1}{\sqrt{2}}&\frac{1}{\sqrt{2}}\\
0&0&-\frac{1}{\sqrt{2}}&\frac{1}{\sqrt{2}}
\end{array}\right)
\left(\begin{array}{cccc}
\ i \ &0&0&0\\
0&\ 1 \ &0&0\\
0&0&\ i \ &0\\
0&0&0&\ 1 \
\end{array}\right)
\left(\begin{array}{cccc}
c_{\gamma_-}&0&s_{\gamma_-}&0\\
0&c_{\gamma_+}&0&s_{\gamma_+}\\
-s_{\gamma_-}&0&c_{\gamma_-}&0\\
0&-s_{\gamma_+}&0&c_{\gamma_+}
\end{array}\right),
\eeq
such that $\mathcal{O}^t\mathcal{M}\mathcal{O} = 
\text{diag}(m_{1_-},m_{1_+},m_{2_-},m_{2_+})$ with 
\beq
\tan(2\gamma_-) &=& \frac{-\sin(2\alpha)\,M}{m_2-m_1+M\,\cos(2\alpha)},\\
\tan(2\gamma_+) &=& \frac{\sin(2\alpha)\,M}{m_2-m_1-M\,\cos(2\alpha)}, 
\nnmb
\eeq
and mass eigenvalues
\beq
m_{1,2_{-}} = \frac{1}{2}\left[
m_1+m_2-M
\mp \sqrt{(m_2-m_1)^2+2(m_2-m_1)M\cos(2\alpha)+M^2}
\right],
\\
m_{1,2_{+}} = 
\frac{1}{2}\left[
m_1+m_2+M
\mp \sqrt{(m_2-m_1)^2-2(m_2-m_1)M\cos(2\alpha)+M^2}
\right].
\nnmb
\eeq
For the specific and technically natural scenario of $M \ll m_2-m_1$ 
that we focus on here, we have
\beq
\Delta m \equiv m_{1_+}-m_{1_-} \simeq 2\,M\,c_{\alpha}^2 \ ,\qquad
m_{2_+}-m_{2_-} \simeq 2\,M\,s_{\alpha}^2 \ ,
\eeq
together with 
\beq
\gamma_\mp \simeq \mp s_\alpha c_{\alpha}\,\frac{M}{m_2-m_1} \ ,
\eeq
corresponding to parametrically small Majorana mass splittings of 
would-be Dirac fermions.

 Adding the additional mixing due to the Majorana mass to the interactions
of the minimal theory, the terms relevant to our dark matter discussion are
\beq
-\lag & \ \supset \ & 
-\frac{\lambda}{2\sqrt{2}}\,\sin(2\alpha+2\gamma_-)\,
h\;\bar{\psi}_{1_-}\psi_{1_-} \\
&&-i\,\bar{\psi}_{1_-}\gamma^{\mu}\psi_{1_+}\bigg(
\cos(\gamma_+-\gamma_-)\,g_x\,X_\mu\;
+\left[\cos(\gamma_+-\gamma_-)-\cos(2\alpha+\gamma_++\gamma_-)\right]
\,\frac{\bar{g}}{4}\,Z_{\mu}\bigg)
\ ,
\nnmb 
\label{eq:majint}
\eeq
where $\psi_{1_\mp} = (\chi_{1_\mp},\bar{\chi}_{1_\mp})^t$ are the 4-component
Majorana fermions constructed from $\chi_{1_-}$ and $\chi_{1_+}$,
and $\bar{g}=\sqrt{g^2+{g'}^2}$.
Here, we have only included Higgs terms involving $\psi_{1_-}$ alone
and dark vector couplings connecting $\psi_{1_-}$ to $\psi_{1_+}$.
Importantly, we note that there are no diagonal vector couplings
between $\psi_{\onem}$ and the vector bosons.

\subsection{Implications for Dark Matter}

Dark matter freezeout with a parametrically small Majorana mass term proceeds 
as in the minimal Dirac theory provided $\Delta m \ll T_{fo} \sim m_1/25$, 
with approximately equal densities of $\psi_{\onem}$ and $\psi_{\onep}$ produced. 
The heavier $\psi_{\onep}$ will then de-excite through decay 
via $\psi_{\onep}\to \psi_{\onem}+\{Z^*,X^*\}$ or by scattering with 
the cosmological bath~\cite{Finkbeiner:2009mi,CarrilloGonzalez:2021lxm}. 
In the limit that 
$m_1,\,m_X \gg \Delta m \gg m_f$, the partial width to SM fermions
$\psi_{\onep}\to \psi_{\onem}+f\bar{f}$ is
\beq
\Gamma_f & \simeq &  \frac{1}{60\pi^3}\left(a_f^2+b_f^2\right)(\Delta m)^5 \\
&\simeq& (9.0\times 10^{-7}\,\text{s})^{-1}\,\frac{(a_f^2+b_f^2)}{G_F^2}\lrf{\Delta m}{100\,\mev}^5,
\nnmb
\eeq
while the thermally averaged scattering cross section for 
$\psi_{\onep}+f\to \psi_{\onem}+f$ de-excitation with $m_f \ll T \ll \Delta m$ is
\beq
\langle\sigma v\rangle \ \simeq \ \frac{1}{2\pi}\,
\left(a_f^2+b_f^2\right)(\Delta m)^2 \ ,
\eeq
with
\beq
a_f \simeq -\frac{4\pi\,\epsilon\,\sqrt{\alpha\,\alpha_x}}{m_x^2}
+ \sqrt{2}\,G_F\,s_{\alpha}^2\,(t^3_f-2\,Q_f\,s_W^2) \ , \qquad
b_f \simeq -\sqrt{2}\,G_F\,s_{\alpha}^2\,t^3_f \ ,
\eeq
where $t^3_f$ is the weak isospin of the left-handed fermion component
and $Q_f$ is the fermion electric charge. We find that the decays are typically 
rapid relative to the start of primordial nucleosynthesis 
($\tau \lesssim 0.1\,\text{s}$) for $\Delta m$ larger than a few tens of MeV,
and that de-excitation by scattering further depletes the heavier state.
The resulting cosmological density of $\psi_{1_-}$ therefore matches that
of $\psi_1$ computed previously in the minimal theory for appropriate
values of $M \ll m_1$. 

In contrast to freezeout, direct detection of $\psi_{1_-}$ dark matter
is impacted very significantly by the Majorana mass splitting.
This splitting leads to exclusively off-diagonal $\psi_{1_-}$ gauge boson
interactions involving the heavier fermions, thereby eliminating elastic
vector exchange contributions to spin-independent~(SI) scattering on nuclei.
Inelastic scattering through vector exchange is still possible, but the rates
for this are highly suppressed or non-existent for a standard local DM halo
velocity distribution and $\Delta m \gtrsim 200$--$500\,\kev$,
depending on the target~\cite{Bramante:2016rdh,Song:2021yar}.
The leading contribution to SI scattering then comes from 
Higgs boson exchange as described by the interactions of Eq.~\eqref{eq:majint}.
This coupling generates $f_p$ and $f_n$ coefficients of
\beq
f_p = -\tilde{d}_p\left(\frac{2}{9}+\frac{7}{9}\sum_q f_q^p\right) \ ,
\qquad
f_n = -\tilde{d}_n\left(\frac{2}{9}+\frac{7}{9}\sum_q f_q^n\right) \ ,
\eeq
with
\beq
\tilde{d}_p = \tilde{d}_n = -\frac{m_n}{2v}
\frac{\lambda\,\sin(2\alpha+2\gamma_-)}{m_h^2}
\ .
\eeq
The effective SI cross section is given by Eq.~\eqref{eq:sigsi}
with these coefficients.

\begin{figure}[ttt]
 \begin{center}
   \includegraphics[width = 0.47\textwidth]{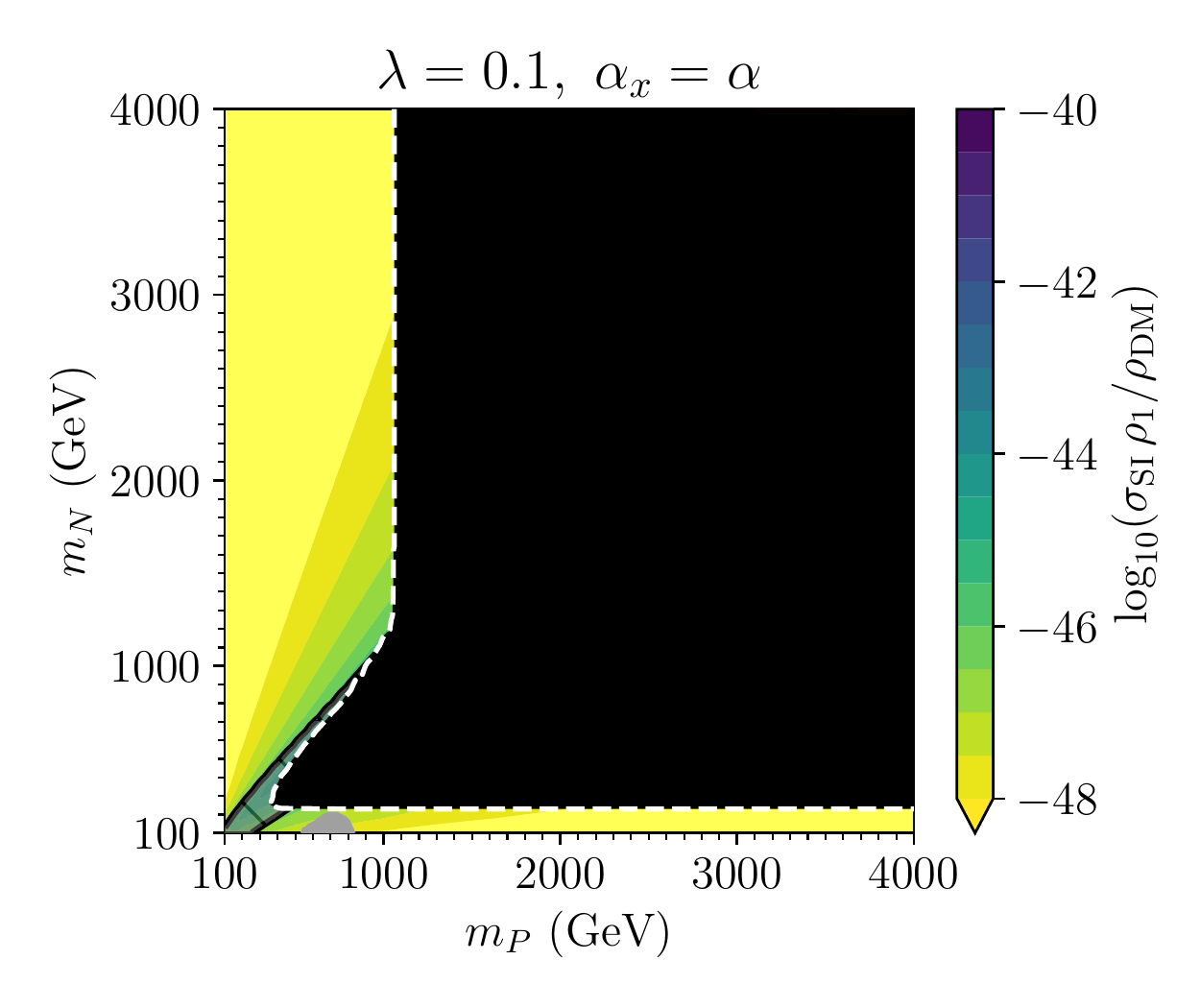}
   \includegraphics[width = 0.47\textwidth]{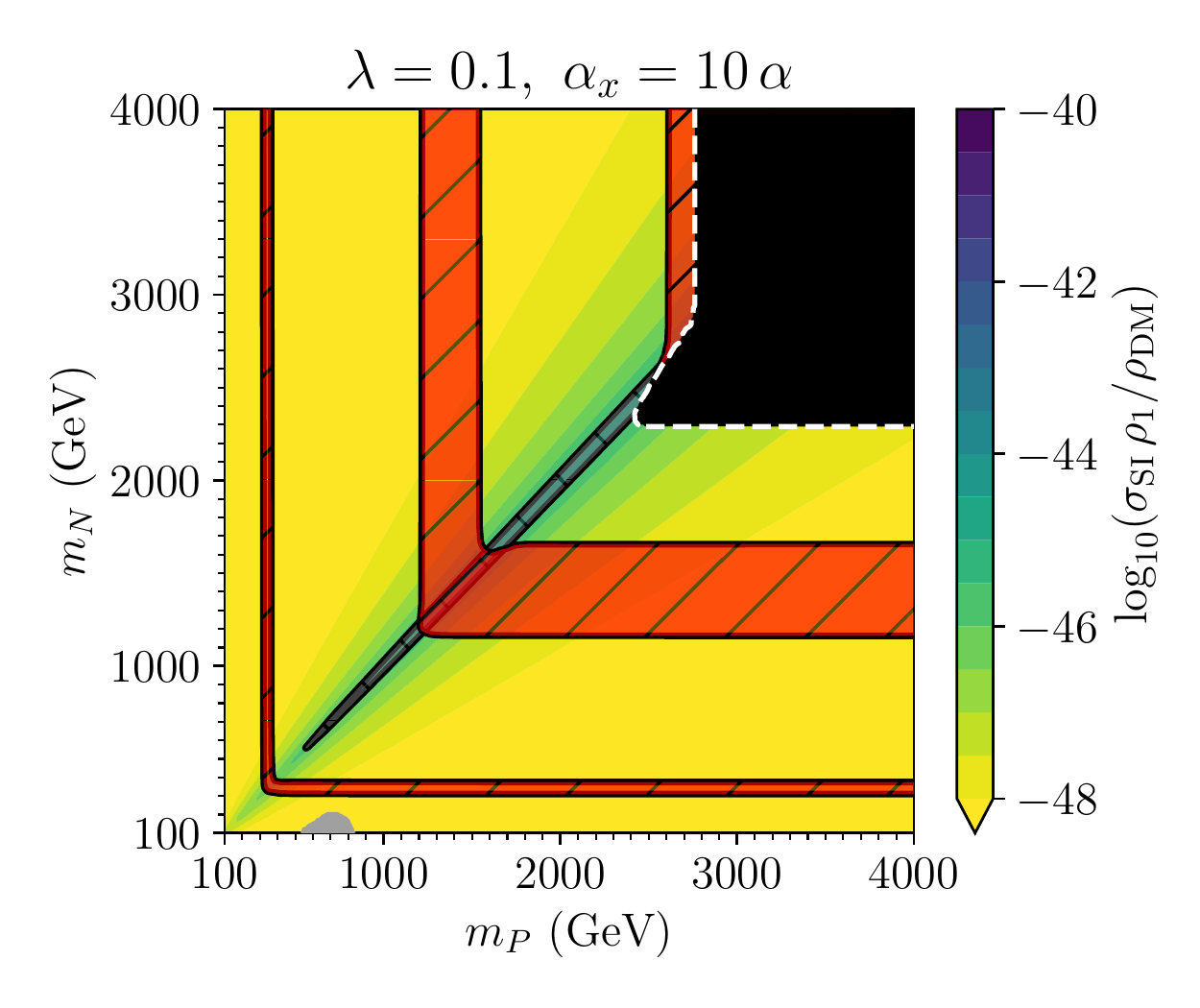}\\
   \includegraphics[width = 0.47\textwidth]{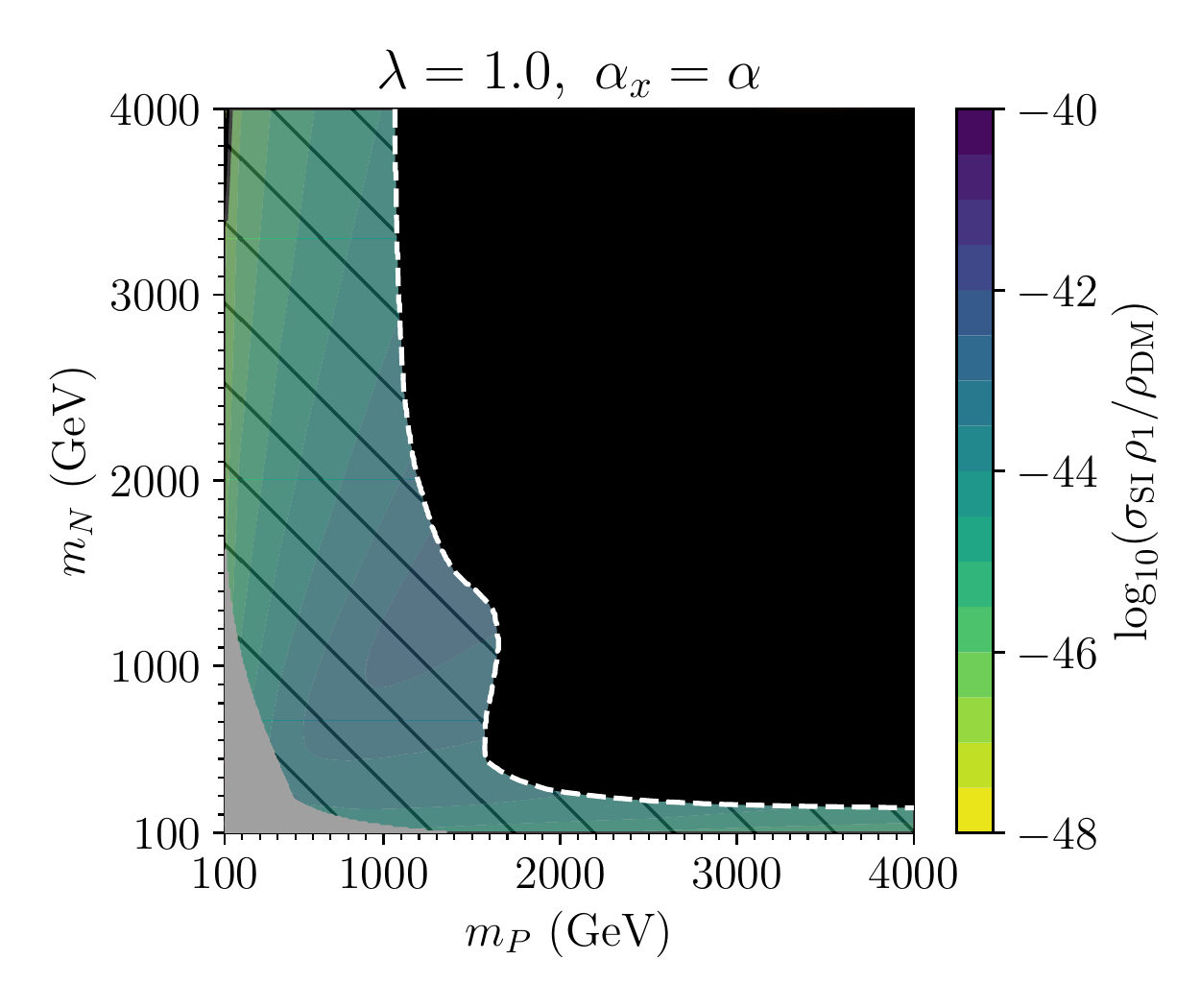}
   \includegraphics[width = 0.47\textwidth]{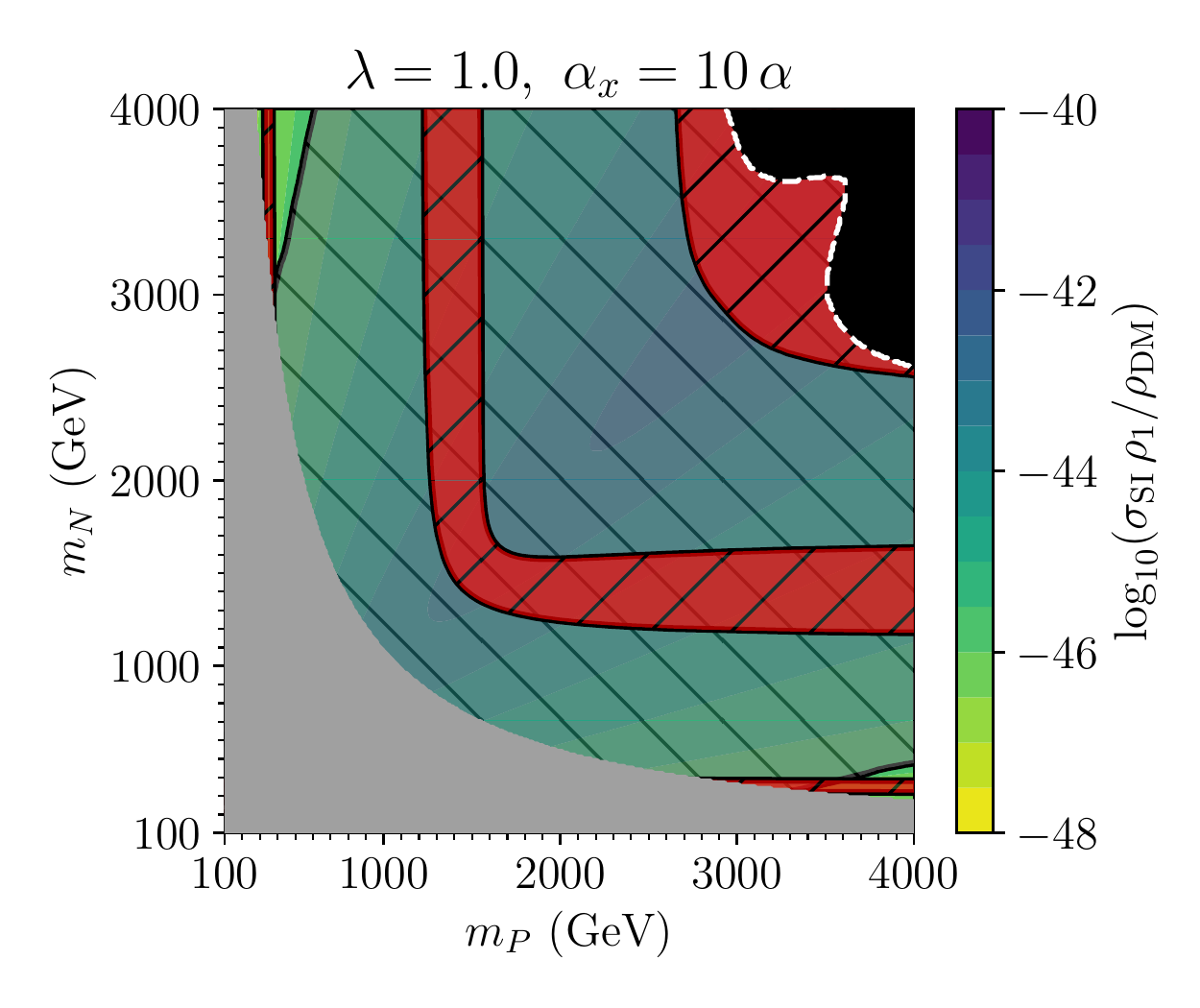}
 \end{center}
\vspace{-0.5cm}
 \caption{Contours of the density-weighted spin-independent
per nucleon cross section $\sigma_\text{SI}\,\rho_1/\rho_\text{DM}$ 
for the Majorana extended theory
in the $m_P$--$m_N$ plane for $\lambda = 0.1,\,1.0$~(top and bottom) 
and $\alpha_x = \alpha,\,10\,\alpha$~(left and right).
The downward hatched regions are excluded by DM direct detection
searches, while the red upward hatched regions are ruled out by
indirect detection tests. Also, the grey shaded regions show the 
combined exclusions from direct searches, 
and the black regions indicate where $\psi_{1_-}$ produces 
too much thermal dark matter.}
 \label{fig:dd2}
 \end{figure}

In Fig.~\ref{fig:dd2} we show the density-weighted SI per-nucleon
cross section, $\sigma_\text{SI}\,\rho_1/\rho_\text{DM}$ in the 
the extended theory of connector fermions with a Majorana mass. We assume 
that $m_2-m_1 \gg \Delta m\gg 500\,\kev$ over the entire $m_P$--$m_N$ plane 
and we set $\lambda=0.1$~(top), $\lambda = 1.0$~(bottom), 
and $\alpha_x=\alpha$~(left) and $\alpha_x = 10\,\alpha$~(right). 
As before, the downward hatched regions show the exclusions 
from DM direct detection experiments, the solid black regions
indicate where the $\psi_{1_-}$ relic density exceeds the observed DM density,
and the grey regions denote exclusions from direct searches.
As expected, the bounds from DM direct detection are significantly impacted
by the Majorana mass. Even so, the remaining nucleon scattering mediated
by Higgs exchange is significant and largely excluded by experiment unless the 
mixing between the singlet and doublet components is small. This can be
achieved for $\lambda \ll 1$, $m_P \ll m_N$, or $m_N \ll m_P$, and is
directly analagous to the suppression of direct detection scattering
for Higgsino-like dark matter in supersymmetry.

Beyond the exclusions shown in Fig.~\ref{fig:dd2} from DM direct detection
and the thermal overproduction of $\psi_{1_-}$, further restrictions
on the parameter space can be derived from indirect DM searches such as gamma rays
emitted from satellite galaxies~\cite{Hess:2021cdp} 
and the inner galaxy~\cite{Fermi-LAT:2017opo,HESS:2022ygk}, 
and distortions in the power spectra of the cosmic microwave 
background~(CMB)~\cite{Chen:2003gz,Padmanabhan:2005es,Slatyer:2009yq}. 
In particular, for larger $\alpha_x \sim 10\,\alpha$ and $m_1 \gg m_x$ 
there can be a significant enhancement of the $\psi_{1_-}$ 
annihilation cross section at late times relative to freezeout due to 
the Sommerfeld effect and bound state 
formation~\cite{Arkani-Hamed:2008hhe,Pospelov:2008jd}. 
To estimate the enhancement relative to perturbative $s$-wave annihilation, 
we follow Refs.~\cite{Cassel:2009wt,Slatyer:2009vg} 
and model the non-relativistic potential 
from dark vector boson exchange with a Hulth\`en potential, 
which has been shown to give a good approximation
for appropriate choices of parameters. Note that since we
focus on parametrically small mass splittings, $\Delta m \ll m_x,\,m_1$,
the Majorana mass is not expected to impact this result 
meaningfully~\cite{Slatyer:2009vg}. 

We compare the enhanced cross sections for late-time $\psi_{1_-}$ annihilation
computed this way to the bound on DM annihilation obtained from CMB observations 
by Planck~\cite{Planck:2018vyg}:
$p_{ann}(\text{DM}) \equiv f_{ann}\,\langle \sigma v\rangle/m_{DM} 
< p_{ann}^{lim} = 3.2\times 10^{-28}\,\text{cm}^3\,\text{s}^{-1}\gev^{-1}$,
where $f_{ann}$ is an efficiency factor that depends on the annihilation products
and $\langle \sigma v\rangle$ is evaluated during recombination.
This CMB bound is comparable to (but usually slightly weaker than) those
from obtained gamma ray measurements~\cite{Hess:2021cdp}, 
but it is also avoids potentially large astrophysical uncertainties. 
In evaluating $p_{ann}(\psi_{1_-})$ for $\psi_{1_-}$ annihilation,
we use a DM velocity of $v \sim \sqrt{0.3\,\text{eV}/m_1}$ 
to estimate the non-perturbative enhancement factor and
we fix $f_{ann}=0.2$~\cite{Cline:2013fm,Slatyer:2015jla}.
We also rescale by the square of the fractional density of $\psi_{1_-}$
relative to dark matter and thus impose the requirement
\beq
p_{ann}(\psi_{1_-}) < p_{ann}^{lim}\lrf{\rho_{\rm DM}}{\rho_1}^2 \ .
\eeq
The exclusions derived in this way are shown by the
red upward hatched regions in Fig.~\ref{fig:dd2}.
These appear only for larger dark gauge couplings $\alpha_x = 10\,\alpha$,
and they exhibit a band structure corresponding to strong enhancement 
in specific regions through the formation of bound states. The locations
of these exclusion bands depend on the mediator mass relative to the DM mass,
but lie in a saturation regime that is largely insensitive to the DM velocity 
used in the calculuation provided it is small enough.

\subsection{Implications for Direct Searches}

  The Majorana mass term of Eq.~\eqref{eq:mmaj} requires the introduction
of a new dark Higgs to the theory and modifies the spectrum relative
to the minimal theory considered in Sec.~\ref{sec:min}. We focus here on
parametrically small mass splittings, $\Delta m \sim M = y_N\eta \ll m_x,\,m_1$,
since they are sufficient to moderate dark matter direct detection.
Given that $\eta = m_x/g_x$, this corresponds to the limit of $y_N \ll g_x$.
With this hierarchy of couplings, the effect of the new operator
need not significantly alter the non-dark matter phenomenology of the theory.

 Starting with the dark Higgs itself, the new physical scalar $\varphi$ obtained
from $\Phi \to (\eta + \varphi/\sqrt{2})$ has efficient
decays $\varphi \to X+X^{(*)}$ provided its mass is larger than the vector's,
$m_\varphi \geq m_x$. We note, however, that for $m_\varphi < m_x$ the 
dark Higgs can be long-lived and potentially problematic for 
cosmology~\cite{Batell:2009yf,Chan:2011aa,Morrissey:2014yma,Berger:2016vxi}.
The coupling of Eq.~\eqref{eq:mmaj} will also induce a Higgs portal
operator $\lambda_{H\Phi}\,H^{\dagger}H\,\Phi^{\dagger}\Phi$ with
$\lambda_{H\Phi} \sim \lambda^2y_N^2s^2_{2\alpha}/(4\pi)^2 \lesssim (\Delta m/m_x)^2\alpha_x\lambda^2/4\pi$ through fermion loops, which is safely small
for the $\Delta m/m_x \ll 1$ limit we focus on~\cite{Agrawal:2021dbo}.
Moreover, with $\lambda_N \ll g_x$ the impact of the dark Higgs 
on $\psi_{1_\mp}$ freezeout, annihilation, and direct detection is negligible.

 In the limit of a parametrically small Majorana mass splitting 
among the fermions, the bounds from precision electroweak, 
Higgs decays, and direct collider searches will be essentially 
the same as found previously. As the mass splitting grows,
a potentially interesting effect in collider searches is 
$\psi_{1_+}\to  \psi_{1_-}+X^{(*)}$ at the end of decay cascades.
These could appear in the multi-purpose LHC detectors~\cite{Izaguirre:2015zva}
or in far detectors dedicated to long-lived particles~\cite{Curtin:2018mvb,Berlin:2018jbm,FASER:2018eoc}.

\section{Decays through Lepton Mixing
\label{sec:lmix}}

A second approach to addressing the overabundance of relic portal fermions 
is to enable them to decay quickly enough to eliminate them 
as cosmological relics.
Such decays can occur at the renormalizable level if there
exists a dark Higgs field $\phi$ with dark charge $q_\phi=q_x$ that develops
a VEV, $\langle\phi\rangle = \eta$. Such a field allows the $P$ doublet 
to mix with the SM lepton doublets through the operator
\beq
-\lag \ \supset \ \lambda_{a}\,\phi\,\overline{P}_R\,L_{La} + (h.c.) \ ,
\label{eq:lmix}
\eeq
where $L_{La} = (1,2,-1/2;0)$ is the SM lepton doublet with flavor 
$a=e,\,\mu,\,\tau$. The cost of these interactions
is that they can induce lepton flavour violation~(LFV). In this section we
show that the couplings of Eq.~\eqref{eq:lmix} can allow $\psi_1$ to
decay on cosmologically short timescales while not violating current
bounds on LFV, even without imposing any particular flavour structure 
on the operators.

\subsection{Charged Lepton Mixing}

The interactions of Eq.~\eqref{eq:lmix} mix the SM charged leptons
with the $P^-$ fermion. Working in a basis where the lepton Yukawa couplings
are diagonal with $m_a = Y_a\langle H^0\rangle$, 
the charged lepton mass terms take the form
\beq
-\lag \ \supset \ \bar{\psi}_R^-\,M_{\mp}\,\psi_L^- + h.c. \ ,
\eeq
where $\psi_R^- = (P^-_R,e_{R\,a})^t$, $\psi_L^- = (P^-_L,e_{L\,a})^t$,
and
\beq
M_{\mp} = \left(
\begin{array}{ccc}
m_P&\lambda_e\eta&\lambda_\mu\eta\\
0&m_e&0\\
0&0&m_{\mu}
\end{array}
\right) \ .
\eeq
Note that we have only written two SM generations here for brevity.
The generalization to three generations is straightforward.

To diagonalize the charged mass matrix, we make the field transformations
\beq
\psi_L^- = U^{\dagger}\psi_L^{-\,\prime} \ , \qquad 
\psi_R = V^{\dagger}\psi_R^{-\,\prime} \ ,
\eeq
with unitary matrices $U$ and $V$ such that 
$VMU^{\dagger} = {diag}(m_P',m_e',m_\mu')$. This implies that 
$VM_{\mp}M_{\mp}^{\dagger}V^{\dagger} = UM_{\mp}^{\dagger}M_{\mp}U^{\dagger}$.
To linear order in the small quantities 
$m_a/m_P,\,\lambda_a\eta/m_P \ll 1$, 
\beq
U^{\dagger} &=& \left(
\begin{array}{ccc}
1&-\lambda_e\eta/m_P&-\lambda_\mu\eta/m_P\\
\lambda_e\eta/m_P&1&0\\
\lambda_\mu\eta/m_P&0&1
\end{array}
\right) + \mathcal{O}(\varepsilon^3) \ ,
\label{eq:umix} \\
&&\nnmb\\
V^{\dagger} &=& 
\left(
\begin{array}{ccc}
1&-(\lambda_e\eta/m_P)^2&-(\lambda_\mu\eta/m_P)^2\\
(\lambda_e\eta/m_P)^2&1&0\\
(\lambda_\mu\eta/m_P)^2&0&1
\end{array}
\right) + \mathcal{O}(\varepsilon^3) 
\label{eq:vmix}
\ .
\eeq
Furthermore, the mass matrix has two zero eigenvalues for $m_e,\,m_\mu \to 0$
implying $m_a' = m_a\,(1+\mathcal{O}(\varepsilon^2))$
where $\varepsilon$ denotes either $m_a/m_P,\,\lambda_a\eta/m_P \ll 1$, 
and thus the SM lepton masses remain proportional to their Yukawa couplings 
to the SM Higgs. The mass of the heavy state $P^{-\prime}$ 
also remains equal to $m_P$ at this order.

\subsection{Neutral Lepton Mixing}

The mass matrix for the neutral leptons is
\beq
-\lag \ \supset \ \bar{\psi}_R^0M_0\psi_L^0 + h.c
\eeq
where $\psi_R^0 = (N^0_R,P^0_R)^t$, 
$\psi^0_L = (N^0_L,P^0_L,\nu_{L\,e},\nu_{L\,\mu})^t$, and
\beq
M_0 = \left(\begin{array}{cccc}
m_N&\lambda v&0&0\\
\lambda v&m_P&\lambda_e\eta&\lambda_\mu\eta
\end{array}\right) \ ,
\eeq
where again we show only the first two generations for brevity.

To find the mass eigenstates, we take
\beq
\psi_R^0 = \mathcal{O}^{\dagger}B^{\dagger}\psi_R^{0\,\prime} \ ,\qquad
\psi_L^0 = \left(\begin{array}{cc}\mathcal{O}^{\dagger}&0\\0&\mathbb{I}\end{array}\right)A^{\dagger}\psi_L^{0\,\prime} \ ,
\eeq
where all matrices are unitary, and
\beq
\mathcal{O}^{\dagger} = \left(\begin{array}{cc}
c_\alpha&s_\alpha\\
-s_\alpha&c_\alpha
\end{array}\right) 
\eeq
corresponds to the mixing angles defined in Eq.~\eqref{eq:psimix}.
This choice implies that as $\eta \to 0$ 
the matrices $A^{\dagger}$ and $B^{\dagger}$ vanish
and the non-zero masses are $m_1$ and $m_2$ as before. 
These matrices can be found from the eigenvectors of
$(B\mathcal{O})M_0M_0^{\dagger}(\mathcal{O}^{\dagger}\!B^{\dagger})$ 
for $B^{\dagger}$ and
$(A\widetilde{\mathcal{O}})M_0^{\dagger}M_0(\widetilde{\mathcal{O}}^{\dagger}\!A^{\dagger})$ 
for $A^{\dagger}$.
Expanding in the small quantities $\varepsilon = \lambda_a\eta/m_{1,2}$, we find
$B^{\dagger} = \mathbb{I}+\mathcal{O}(\varepsilon^2)$ and
\beq
A^{\dagger} = \left(\begin{array}{cc}
\mathbb{I}&D^{\dagger}\\
-D&\mathbb{I}
\end{array}\right) + \mathcal{O}(\varepsilon^3) \ ,
\qquad
D^{\dagger} = \left(\begin{array}{cc}
s_{\alpha}\lambda_e\eta/m_1~&s_{\alpha}\lambda_\mu\eta/m_1\\
-c_{\alpha}\lambda_e\eta/m_2~&-c_{\alpha}\lambda_\mu\eta/m_2
\end{array}\right) \ .
\eeq
Based on the structure of these matrices, we are guaranteed to have two
massless eigenvalues that we identify with the SM-like neutrinos.
This implies that the new interaction of Eq.~\eqref{eq:lmix} 
does not generate neutrino masses on its own and corrections to 
neutrino masses (necessarily from other sources) from the mixing 
are proportional to those masses.
The non-zero mass eigenvalues remain $m_1$ and $m_2$ up to fractional 
corrections of order $\mathcal{O}(\varepsilon^2)$.

\subsection{Interactions and Heavy Fermion Decay}

In Appendix~\ref{sec:appa}, we collect the interactions between the
new heavy fermions and the scalar and vector bosons of the SM,
as well as the new lepton flavour mixing interactions between the SM fermions
induced by the interaction of Eq.~\eqref{eq:lmix}. Expanding in powers 
of $\varepsilon \sim \{m_a,\lambda_a\eta\}/m_{1,2} \ll 1$, 
heavy-light lepton interactions arise at linear order in $\varepsilon$
while LFV light-light lepton interactions are quadratic in $\varepsilon$.
The lone exception to this comes from the interactions with the dark Higgs boson
$\varphi$, obtained from $\phi \to (\eta+\varphi/\sqrt{2})$ in unitary gauge,
which produce heavy-light fermion couplings at zeroeth order in $\varepsilon$.

Based on the structure of these couplings and counting powers of $\varepsilon$, 
the dominant $\psi_1$ decay channels would appear to be 
$\psi_1 \to \nu_{L a}+\varphi$, with width
\beq
\Gamma(\psi_1\to \nu_{L a}\,\varphi) = 
\frac{\lambda_a^2s_{\alpha}^2}{64\pi}\,m_1\left[1-\lrf{m_{\varphi}}{m_1}^2\right] \ .
\label{eq:lpfdec}
\eeq
However, for $m_1\gg m_x$ the decays $\psi_1\to \nu_{La} +X^{\mu}$ receive
a longitudinal enhancement,
\beq
\Gamma(\psi_1\to \nu_{L a}\,X) = 
\frac{\lambda_a^2s_{\alpha}^2}{64\pi}\,m_1\left[1-\lrf{m_x}{m_1}^2\right]
\left[1+2\lrf{m_x}{m_1}^2\right] \ .
\label{eq:lpxdec}
\eeq
In the limit $m_1 \gg m_\varphi,\,m_x$, the sum of these widths reproduces
the width for $\psi_1\to \nu_{L a}+\phi$ in the $U(1)_x$-unbroken theory,
as expected from the Nambu-Goldstone equivalence 
theorem~~\cite{Cornwall:1974km,Vayonakis:1976vz,Chanowitz:1985hj}.
Relative to the decay channels involving $\varphi$ or $X^{\mu}$,
all other modes are suppressed by factors of at least 
$(m_x/m_V)^2,\,\varepsilon^2 \ll 1$.

Using the decay widths above, the total lifetime of $\psi_1$ is 
\beq
\tau \ \simeq \ 
(6.61\times 10^{-8}\,\text{s})
\lrf{10^{-9}}{\lambda_as_{\alpha}}^2\lrf{\tev}{m_1} \ .
\label{eq:taupsi}
\eeq
As long as the couplings are not exceedingly small,
$\lambda_as_{\alpha} \gtrsim 10^{-12}$, these decays occur
before primordial nucleosynthesis and neutrino decoupling, and will generally
be safe from cosmological bounds~\cite{Kawasaki:2004qu,Jedamzik:2006xz}.  
Of course, this also eliminates $\psi_1$ as a relic particle and removes
the bounds from direct detection since $\rho_1\to 0$ today.

\subsection{Lepton Flavor Mixing}

The couplings collected in Appendix~\ref{sec:appa} for this scenario
give rise to lepton flavor violation~(LFV) and modify
leptonic anomalous magnetic moments 
$\Delta a_\ell \equiv a_\ell-a_\ell^{\text{SM}}$. 
We compute here the new effects of the mixing from Eq.~\eqref{eq:lmix}
on $\Delta a_\ell$ and LFV observables~\cite{Hisano:1995cp,Calibbi:2017uvl,Calibbi:2018rzv,Crivellin:2020ebi}. 

In analogy to the decay calculation above, and again counting powers of
$\varepsilon \sim \lambda_a\eta/m_P\ll 1$, $(m_x/m_Z)^2 \ll 1$,
the dominant new contributions to $\Delta a_\ell$ and LFV come from loop 
diagrams with the dark Higgs $\varphi$ or the dark vector $X^\mu$ together
with internal lines involving the heavy $P^-$ fermion. 
For $m_P\gg m_\varphi,\,m_x$ the two contributions are approximately equal
and sum to the result obtained by calculating in the $U(1)_x$-unbroken theory
provided $m_\phi \ll m_P$. 

To see this explicitly, consider the leading contribution to the amplitude 
for $\ell_a(p)\to \ell_b(p')+\gamma(q)$ in the unbroken theory, corresponding
to vertex and external leg diagrams with loops containing 
$\phi$ and $P^-$. We find
\beq
-i\mathcal{M}_{ab} \simeq  
\frac{i\,e\,\lambda_a\lambda_b^*}{32\,\pi^2}\,f(m_P,m_\phi)\,
\bar{u}'i\sigma^{\mu\nu}q_{\nu}(m_bP_L+m_bP_R)u\,F_{\mu\nu} + (\ldots) \ ,
\label{eq:lfvamp}
\eeq
where the omitted remainder is not relevant for $\Delta a_\mu$ 
or $\ell_a \to \ell_b+\gamma$ transitions,
and the loop function $f(m_P,m_\phi)$ is
\beq
f(m_P,m_\phi) &=& \int_0^1\!dz\frac{z(1-z)^2}{(1-z)m_P^2+zm_{\phi}^2}
\label{eq:lfvloop}\\ 
&=&
\frac{1}{6m_{\phi}^2}\left[\frac{2+3r-6r^2+r^3+6r\ln r}{(1-r^4)}\right]\nnmb\\
&\to& \frac{1}{6m_P^2}~~~~~~~~~~(r \to \infty) \ ,
\nnmb
\eeq
with $r=m_P^2/m_{\phi}^2$. Note that we have self-consistently neglected 
lepton masses $m_a\ll m_P$ beyond the leading non-trivial order 
and that our result matches 
Refs.~\cite{Hisano:1995cp,Calibbi:2017uvl,Calibbi:2018rzv}.
The impact of $m_P$ and $m_\phi$ is seen in the loop function,
and Eq.~\eqref{eq:lfvloop} shows that the result is independent
of $m_\phi$ for $m_P^2\gg m_\phi^2$ as claimed. We find the
same leading result in the $U(1)_x$-broken theory for $m_\varphi,\,m_x \ll m_P$.

 Using Eq.~\eqref{eq:lfvamp}, we can extract the dominant contributions
to $\Delta a_{\ell}$ as well as rates for $\ell_a\to \ell_b+\gamma$.
For the first, we have
\beq
\Delta a_{\ell_a} = +\frac{\lambda_a^2}{96\pi^2}\lrf{m_{a}}{m_P}^2 \ .
\eeq
In turn, this can be related to the $\ell_a \to \ell_b+\gamma$ LFV rates 
by~\cite{Calibbi:2018rzv}
\beq
\text{BR}(\mu\to e\gamma) &=& 
\frac{12\pi^3}{m_\mu^4}\frac{\alpha}{G_F^2}\lrf{\lambda_e}{\lambda_\mu}^2
\times (\Delta a_{\mu})^2,
\\
\text{BR}(\tau\to \mu\gamma) &=& 
\frac{12\pi^3}{m_\mu^4}\frac{\alpha}{G_F^2}\lrf{\lambda_\tau}{\lambda_\mu}^2
\times (\Delta a_{\mu})^2
\times \text{BR}(\tau\to \mu\nu\bar{\nu}) \ ,
\eeq
with $\text{BR}(\tau\to\mu\nu\bar{\nu}) = 0.174$~\cite{ParticleDataGroup:2020ssz}.

In this theory the rates for $\ell_a \to \ell_b+\gamma$ also give a good
proxy for other LFV observables. 
As an example, the amplitude for $\ell_a \to 3\ell_b$ has dipole and non-dipole 
contributions involving off-shell intermediate vector bosons that scale 
as $e\lambda_a\lambda_b$, as well as box contributions that 
go like $\lambda_a\lambda_b^3$. Since we focus on $\lambda_a \ll e,\,g_x$,
the box diagrams are subleading relative to the off-shell vector contributions
which have the same parametric dependence on the couplings as 
$\ell_a\to \ell_b+\gamma$. Put together, we expect 
$\text{BR}(\ell_a\to 3\ell_b) \lesssim \alpha\,\text{BR}(\ell_a\to\ell_b\gamma)$~\cite{Hisano:1995cp,Calibbi:2017uvl,Toma:2013zsa}.
A similar argument applies to $\mu\,N\to e\,N$ conversion.
We also note that direct dark vector contributions to $\Delta a_{\ell}$
are negligible for the values $m_x=15\,\gev$ 
and $|\epsilon| \lesssim 10^{-3}$ that we focus on~\cite{Pospelov:2008zw}.

\begin{figure}[ttt]
 \begin{center}
         \includegraphics[width = 0.47\textwidth]{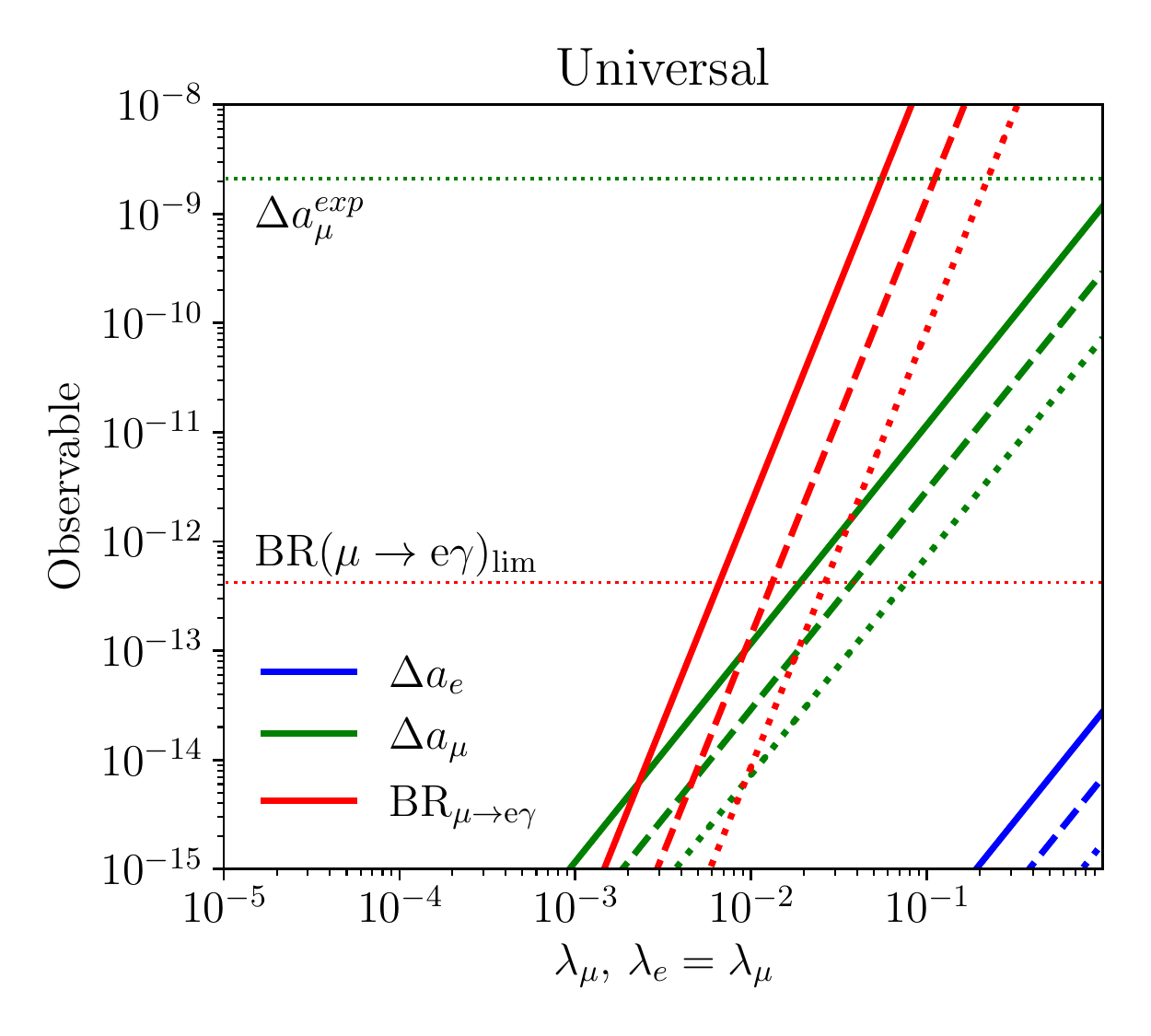}
         \includegraphics[width = 0.47\textwidth]{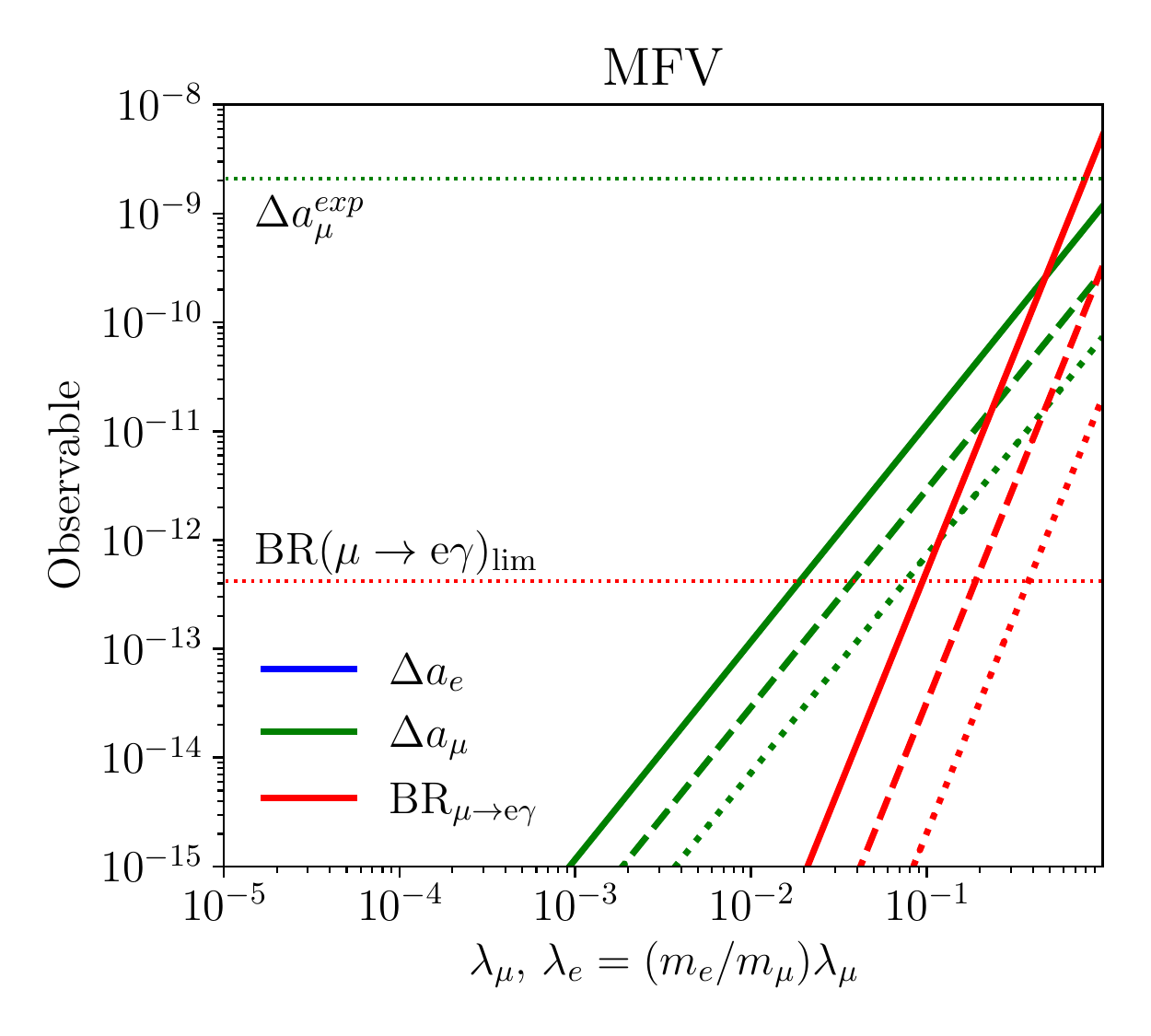}
 \end{center}
\vspace{-0.5cm}
 \caption{Charged lepton observables $\Delta a_{e,\mu}$ 
and $\text{BR}(\mu\to e \gamma)$ in the lepton mixing model as
a function of $\lambda_\mu$ assuming universal~($\lambda_e=\lambda_\mu$)
or MFV-inspired ($\lambda_e=(m_e/m_\mu)\lambda_\mu$) couplings.
For each observable, the solid, dashed, and dotted lines correspond 
to $m_P = 100,\,200,\,400\,\gev$.}
 \label{fig:lfvdiag}
 \end{figure}

In Fig.~\ref{fig:lfvdiag} we show the lepton mixing model predictions for
the observables $\Delta a_e$~(blue), $\Delta a_\mu$~(green),  
and $\text{BR}(\mu\to e\gamma)$~(red) in terms of the coupling 
$\lambda_\mu$ for a universal scenario with $\lambda_e = \lambda_\mu$~(left) and 
a minimal flavour violation~(MFV) inspired scenario 
with $\lambda_e = (m_e/m_\mu)\lambda_\mu$~(right).
For all three observables, the solid lines correspond to $m_P=100\,\gev$,
the dashed lines to $m_P=200\,\gev$, and the dotted lines to $m_P=400\,\gev$. 
The horizontal dotted lines indicate the combined experimental central value 
for $\Delta a_\mu = (2.51\pm 0.59)\times 10^{-9}$~\cite{Muong-2:2006rrc,Muong-2:2021ojo} and the current bound on
$\text{BR}(\mu\to e\gamma) < 4.2\times 10^{-13}$~\cite{MEG:2016leq}.
We see from the figure that LFV provides the most stringent constraint
on the lepton-mixing couplings $\lambda_a$, with $\lambda_\mu \lesssim
5\times 10^{-3}$ sufficent for consistency with the current experimental result
in the scenarios considered. Comparing to Eq.~\eqref{eq:taupsi}, this implies 
that LFV bounds can easily be consistent with mediator fermion decays
well before the era of primordial nucleosynthesis.

\subsection{Other Bounds}

  Lepton mixing in this scenario requires the presence of a new dark Higgs 
scalar $\varphi$. As in the Majorana mass scenario of Sec.~\ref{sec:maj}, 
the dark Higgs can be consistent with observations for 
$m_\varphi > m_x$ provided its Higgs portal coupling $\lambda_{\phi H}$ 
is not too large. Such a coupling will be induced by loops 
with $\lambda_{\phi H} \sim \lambda^2\lambda_a^2/(4\pi)^2$
and are therefore naturally small for $\lambda_a \ll 1$.

  Decays of the lightest connector fermion $\psi_1$ can change the collider 
signatures of the theory relative to the minimal model of Sec.~\ref{sec:min}, 
where it is stable and produces missing energy. 
Collider events in this extension start out just like in the minimal theory, 
with dominant pair production of the connector fermions 
through Drell-Yan processes followed by cascade decays of the heavier states 
down to the lightest $\psi_1$ mode. However, $\psi_1$ now decays
further to the SM through the dominant channels $\psi_1 \to \nu_{a}+\varphi$ 
and $\psi_1\to \nu_a+X$. This yields additional visible energy
from the subsequent $X \to f\bar{f}$ and $\varphi\to XX^{(*)}$ decays.
Similar signals have been studied in supersymmetric dark sector models where
the lightest SM superpartner decays to a lighter dark 
sector~\cite{Strassler:2006im,Strassler:2006qa,Han:2007ae,Arkani-Hamed:2008hhe,Arkani-Hamed:2008kxc,Baumgart:2009tn,Cheung:2009su,Chan:2011aa}.
 
If all the $\lambda_a$ couplings are reasonably small, 
$\lambda_a s_\alpha \lesssim 10^{-10}$,
Eq.~\eqref{eq:taupsi} implies that the $\psi_1$ is effectively
stable with respect to standard collider detectors. 
Collider bounds on the theory are then the same as those discussed 
in Sec.~\ref{sec:min} with the $\psi_1$ producing missing energy. 
However, the slow decays of of this lightest connector fermion could 
potentially be observed at dedicated far detectors~\cite{Curtin:2018mvb,Berlin:2018jbm,FASER:2018eoc}. 

 For larger couplings, $\psi_1$ can decay relatively promptly 
on typical collider timescales and generate additional visible energy 
in the events.
In the limit $m_x \ll m_1$ the dark vector decay products will be at least
moderately boosted and may give rise to lepton 
jets~\cite{Arkani-Hamed:2008hhe,Arkani-Hamed:2008kxc}.
These have been searched for in the context of Higgs boson decays
to dark sector particles~\cite{Falkowski:2010cm,Falkowski:2010gv} 
for both prompt~\cite{ATLAS:2021ldb,CMS:2021pcy} and delayed
dark particle decays~\cite{CMS:2022qej,ATLAS:2022izj}, 
but we do not know of a similar experimental analysis for production 
through connector fermions or supersymmetric cascades. 
Some insight into potential collider sensitivities can be obtained from
LHC searches for heavy vector-like leptons that decay to taus
or tau neutrinos. These produce events with $Z$, $W$, or $h$ bosons
together with additional visible energy from taus, instead of dark photons.
These searches constrain doublet-like vector-like leptons up to masses
approaching $m \gtrsim 1000\,\gev$~\cite{CMS:2022nty,ATLAS:2022vod} 
and suggest that similar bounds could be obtained on $m_P$ in 
our lepton mixing scenario. We defer a detailed collider analysis 
of this scenario to a future work, as well as the impact on the sensitivity 
for intermediate mixing couplings that produce moderately 
long-lived $\psi_1$ fermions.

\section{Conclusions\label{sec:conc}}

  In this work we have investigated the experimental and dark matter implications
of vector-like connector fermions between the SM and a dark sector.
Our focus was on a specific example of a dark $U(1)_x$ gauge sector
together connector fermions with gauge charges $N=(1,1,0;q_x)$ 
and $P=(1,2,-1/2;q_x)$ that connect to the SM Higgs field through
a new Yukawa interaction. Electroweak symmetry breaking and the Higgs
Yukawa leads to mixing between the fermions producing a lightest exotic
state $\psi_1$, while gauge invariance without any further structure 
implies that this new state is stable. 

 Being stable, the $\psi_1$ fermion contributes to the total density
of dark matter assuming a standard cosmological history. 
We have studied the relic density it obtains from thermal
freezeout, as well as the signals it produces in dark matter search experiments.
An acceptable relic density of $\psi_1$ can be obtained provided it is not
too heavy, $m_1 \lesssim 3\,\tev$ for reasonable couplings. However, being
a Dirac fermion with couplings to vector bosons, we find that this relic 
is nearly always ruled out by direct detection searches for dark matter,
even when it only makes up a very small fraction of the total relic density.

To address this challenge faced by the connector fermions, we investigated
two extensions of the minimal theory. In the first, we introduced
a dark Higgs field $\Phi$ with dark charge $q_{\Phi}=-2q_x$ that can give
rise to a gauge-invariant Majorana mass term for $\psi_1$, splitting it
into pseudo-Dirac components $\psi_{1_{\mp}}$. For a range of small mass splittings,
the $\psi_{1_-}$ state makes all the dark matter today and does not couple
diagonally to vector bosons. This eliminates the leading contributions to
nucleon scattering and allows the relic to be consistent with current direct
detection bounds. In the second extension, we added a dark Higgs field $\phi$
with $q_\phi = q_x$ that allows the $P$ fermion to mix with the SM
lepton doublets. This leads to decays of the lightest $\psi_1$ fermion 
to SM states but can also induce charged lepton flavour violation~(LFV).
We showed that the decays can be fast enough to be allowed by cosmology while
also being consistent with tests of LFV.

Even though we have considered a specific example of connector fermions,
similar considerations apply more generally to any new states carrying
both SM and dark gauge charge. The lightest of these will be stable
in the absence of dark symmetry breaking, and accidental flavor symmetries may
lead to additional long-lived or stable states. 
When the dark gauge sector is Abelian, minimal mass terms for the connectors 
require Dirac or complex scalar representations, 
and their gauge charges (dark or electroweak) lead
to vector couplings to nucleons. These tend to produce nucleon scattering
cross sections that are much larger than current limits, implying strong
constraints from direct detection if the connector fermions were created
in the early Universe. Let us emphasize further that the bounds on relic 
connector fermions will typically be even more stringent for other SM 
representations of the connectors.

 It is notable that the solutions to this challenge that we have investigated
both involve spontaneous Higgsing of the dark gauge group. Furthermore,
these solutions are only possible for certain, specific gauge representations
of the connector fermions. This has significant implications for 
the effective field theory describing dark photon interactions with 
the SM beyond the kinetic-mixing portal itself. Specifically, these points 
lead to the conclusion that combining the minimal connector theory with 
a Stueckelberg mechanism to generate the dark photon mass is inconsistent 
with a standard (hot) cosmological history. Thus, the need to associate 
the dark photon mass with the presence of a gauge symmetry informs 
the structure of higher-order operators allowed at 
the weak scale~\cite{Kribs:2022gri,Aebischer:2022wnl} 
as well as the expected spectrum of states in the dark sector.

\acknowledgments{
We thank Mary-Jean Harris for contributions at the beginning of this project.
We also thank Hooman Davoudiasl, David McKeen, John Ng, and Nirmal Raj
for helpful discussions.
This work is supported by the Natural Sciences
and Engineering Research Council of Canada~(NSERC), with DM
supported in part by a Discovery Grant. 
TRIUMF receives federal funding via a contribution agreement 
with the National Research Council of Canada.  
}


\appendix

\section{Appendix: Interactions with Lepton Mixing\label{sec:appa}}

Relevant interactions in the minimal theory without lepton mixing are
given in Ref.~\cite{Lu:2017uur}. Here, we extend these to include
general lepton mixing terms of the form of Eq.~\eqref{eq:lmix}.
The couplings of the minimal theory can also be obtained from 
these results by setting $\lambda_a=0$.

The gauge and mass eigenstates involving the connector fermions are now
$\psi^0_{L\,I} = (N^0_L,P^0_L,\nu_{L\,a})$, $\psi^0_{R\,i} = (N^0_R,P^0_R)$,
$\psi^-_{L\,I} = (P^-_R,e_{L\,a})$, $\psi^-_{R\,I} = (P^-_R,e_{R\,a})$,
where we label $I = i,a$, with $a= e,\mu,\tau$ and $i=(1),2$.
In the expressions below, we show the exact result in terms of the
full mixing matrices as well as the leading non-trivial operators
in powers of $\varepsilon = \lambda_a\eta/m_i\ll 1$.
\bigskip

\noindent\textbf{Photon $(\Gamma = e\gamma^{\mu}A_\mu)$}
\beq
-\lag &\supset&  -\bar{\psi}^-_I\Gamma\psi^-_I\\
&=& -\bar{\psi}^{-\prime}_I\Gamma\psi^{-\prime}_I\nnmb
\eeq

\noindent\textbf{Dark Photon ($\Gamma = g_x\gamma^{\mu}X_{\mu})$}
\beq
-\lag &\supset&
\bar{\psi}^0_{L\,i} \Gamma {\psi}^0_{L\,i}
+ \bar{\psi}^0_{R\,i} \Gamma {\psi}^0_{R\,i}
+ \bar{\psi}^-_{L\,i} \Gamma {\psi}^-_{L\,i}
+ \bar{\psi}^-_{R\,i} \Gamma {\psi}^-_{R\,i}\\
&&\nnmb\\
&=&
\bar{\psi}^{0\prime}_{L\,I} \Gamma (A\widetilde{\mathcal{O}})_{Ii}(\widetilde{\mathcal{O}}^{\dagger}\!A^{\dagger})_{iJ}{\psi}^{0\prime}_{L\,J}
+ \bar{\psi}^{0\prime}_{R\,i}\Gamma {\psi}^{0\prime}_{R\,i}\nnmb\\
&&~~+ \bar{\psi}^{-\prime}_{L\,I}\Gamma U_{Ii}U^{\dagger}_{iJ}{\psi}^{-\prime}_{L\,J}
+ \bar{\psi}^{-\prime}_{R\,i}\Gamma V_{Ii}V^{\dagger}_{iJ}{\psi}^{-\prime}_{R\,i}\nnmb\\
&&\nnmb\\
&\simeq&
\bar{\psi}^{0\prime}_{Li}\Gamma{\psi}^{0\prime}_{Li}
+ \bar{\psi}^{0\prime}_{Ri}\Gamma{\psi}^{0\prime}_{Ri}\nnmb\\
&&
+ \lambda_a\eta\,\big[\big(-\frac{s_\alpha}{m_1}\,\bar{\psi}^{0\prime}_{L1}
+\frac{c_\alpha}{m_2}\,\bar{\psi}^{0\prime}_{L2}\big)\Gamma\nu^\prime_{La}+h.c.\big]
- \lambda_a\lambda_b\eta^2\,\big(\frac{s_{\alpha}^2}{m_1^2}+\frac{c_{\alpha}^2}{m_2^2}\big)\,\bar{\nu}^{\prime}_{La}\Gamma\nu^{\prime}_{Lb}
\nnmb\\
&& + \bar{\psi}^{-\prime}_{Li}\Gamma{\psi}^{-\prime}_{Li}
+ \bar{\psi}^{-\prime}_{Ri}\Gamma{\psi}^{-\prime}_{Ri}
-\left[\bar{\psi}^{-\prime}_{Li}\Gamma\,(\lambda_a\eta/m_P)\,e^{\prime}_{La}+h.c.\right]
+ \bar{e}^{\prime}_{La}\Gamma\,(\lambda_a\lambda_b\eta^2/m_P^2)\,e^{\prime}_{L b}
\nnmb
\eeq

\noindent\textbf{Z Boson ($\Gamma = \frac{g}{c_W}\gamma^{\mu}Z_\mu)$}
\beq
-\lag &\supset& 
%
\frac{1}{2}\bar{\psi}^0_{LI}\Gamma(\delta_{IJ}-\delta_{I1}\delta_{1J})\psi^0_{LI} 
+ \frac{1}{2}\bar{\psi}^0_{Ri}\Gamma\delta_{i2}\delta_{2j}\psi^0_{Ri} 
\label{eq:zcoup}\\
&&
+ (-\frac{1}{2}+s^2_W)\bar{\psi}_{LI}^-\Gamma\delta_{IJ}\psi_{LJ}^-
+ \bar{\psi}^-_{RI}\Gamma(s^2_W\delta_{IJ}-\frac{1}{2}\delta_{I2}\delta_{2J})\psi^-_{RJ} 
\nnmb\\
&&\nnmb\\
&=& \frac{1}{2}\bar{\psi}^\zpr_{LI}\Gamma[\delta_{IJ} - (A\widetilde{O})_{I1}(\widetilde{O}^{\dagger}A^{\dagger})_{1J}]\psi^\zpr_{LJ}
+\frac{1}{2}\bar{\psi}^\zpr_{Ri}\Gamma[(B\mathcal{O})_{i2}(\mathcal{O}^{\dagger}B^\dagger)_{2j}]\psi^\zpr_{Rj}
\nnmb\\
&&
+ (-\frac{1}{2}+s^2_W)\bar{\psi}_{LI}^\mpr\Gamma\,\delta_{IJ}\,\psi_{LJ}^\mpr
+ \bar{\psi}_{RI}^\mpr\Gamma\,(s^2_W\delta_{IJ}-\frac{1}{2}V_{I2}V^{\dagger}_{2J})\,\psi_{RJ}^\mpr 
\nnmb\\
&&\nnmb\\
&\simeq&
\frac{1}{2}[s_{\alpha}^2\bar{\psi}_1^\zpr\Gamma\psi_1^\zpr 
+ c_{\alpha}^2\bar{\psi}_2^\zpr\Gamma\psi_2^\zpr - c_{\alpha}s_{\alpha}
(\bar{\psi}_1^\zpr\Gamma\psi_2^\zpr + h.c.)]
\nnmb\\
&& 
+ (-\frac{1}{2}+s_W^2)P^{\mpr}\Gamma P^{\mpr}
+ (-\frac{1}{2}+s_W^2)e_{La}^{\prime}\Gamma e_{La}^{\prime}
+ s_W^2\bar{e}^{\prime}_{Ra}\Gamma e^{\prime}_{Ra}
\nnmb\\
&&
-\lambda_ac_\alpha s_\alpha\eta\lrf{m_2\!-\!m_1}{m_1m_2}(c_\alpha\bar{\psi}_{1L}^\prime+s_{\alpha}\bar{\psi}_{2L}^\prime)\Gamma\nu^{\prime}_{La}+h.c.
\nnmb\\
&&
-\lambda_a\lambda_b\eta^2s_\alpha^2c_{\alpha}^2\lrf{m_2\!-\!m_1}{m_1m_2}^2\bar{e}^{\prime}_{La}\Gamma e^{\prime}_{Lb}
\ .
\nnmb
\eeq

\noindent\textbf{W Boson ($\Gamma = \frac{g}{\sqrt{2}}\gamma^{\mu}W_\mu)$}
\beq
-\lag &\supset& 
\bar{\psi}_{LI'}^-\Gamma\psi^0_{LI'} + \bar{\psi}^-_{RI'}\Gamma\delta_{I'2}\delta_{2j}\psi^0_{Rj} + h.c.
\label{eq:wcoup}\\
&&\nnmb\\
&=&
\bar{\psi}_{LI'}^{\mpr}\Gamma U_{I'K'}(\mathcal{O}^{\dagger}A^{\dagger})_{K'J}\psi^{\zpr}_{L J}
+ \bar{\psi}^{\mpr}_{RI'}\Gamma V_{I'2}(\mathcal{O}^{\dagger}B^{\dagger})_{2j}\psi^{\zpr}_{Rj} + h.c.
\nnmb\\
&&\nnmb\\
&=&
\bar{P}^{\mpr}\Gamma(-s_\alpha\psi_1^{\zpr}+c_{\alpha}\psi_2^{\zpr})
+ \bar{e}^{\prime}_{La}\Gamma\nu_{La}\nnmb\\
&& 
-\lambda_a\eta\left(\frac{1}{m_P}+\frac{s_{\alpha}^2}{m_1}+\frac{c_{\alpha}^2}{m_2}\right)\bar{P}^{\mpr}_{L}\Gamma\nu_{L a}
- \frac{\lambda_a\eta}{m_P}\bar{e}^{\prime}_{La}\Gamma(-s_{\alpha}\psi_{L1}^{\zpr}+c_{\alpha}\psi_{L2}^{\zpr})
\nnmb\\
&&
+ \frac{\lambda_a\lambda_b\,\eta^2}{m_P}\left(\frac{s_{\alpha}^2}{m_1}+\frac{c_{\alpha}^2}{m_2}\right)\,\bar{e}_{La}^{\prime}\Gamma\nu_{L b}^\prime + h.c. 
\nnmb
\eeq

\noindent\textbf{Higgs Boson ($\Gamma = h/\sqrt{2})$}
\beq
-\lag &\supset& Y_a\,\bar{\psi}_{RI}^-\Gamma\delta_{Ia}\delta_{aJ}\psi_{LJ}
+ \lambda\,\bar{\psi}_{Ri}^0\Gamma\left(\delta_{i1}\delta_{2J}
+\delta_{i2}\delta_{1J}\right)\psi^-_{LJ} + h.c.\\
&&\nnmb\\
&=&
Y_a\,\bar{\psi}_{RI}^-\Gamma V_{Ia}U^{\dagger}_{aJ}\psi_{LJ}
+ \lambda\,\bar{\psi}_{Ri}^0\Gamma\left[
(B\mathcal{O})_{i1}(\mathcal{O}^{\dagger}A^{\dagger})_{2J}
+(B\mathcal{O})_{i2}(\mathcal{O}^{\dagger}A^{\dagger})_{1J}
\right]\psi_{LJ}^\zpr
+ h.c.\nnmb\\
&&\nnmb\\
&\simeq&
Y_a\,\bar{e}'_{Ra}\Gamma e'_{La} + Y_a\,\frac{\lambda_a\eta}{m_P}\,\bar{e}_{Ra}'\Gamma P_L^{\mpr}
\nnmb\\
&& 
\lambda\,s_{2\alpha}\left(-\bar{\psi}_{1R}^\zpr\Gamma\psi^{\zpr}_{1L}
+ \bar{\psi}_{2R}^\zpr\Gamma\psi^{\zpr}_{2L}\right)
+ \lambda\,c_{2\alpha}\left(\bar{\psi}_{1R}^\zpr\Gamma\psi^{\zpr}_{2L} + \bar{\psi}_{2R}^\zpr\Gamma\psi^{\zpr}_{1L}\right)\nnmb\\
&&
-\lambda\,\lambda_a\eta\left(s_{2\alpha}\frac{s_{\alpha}}{m_1}+ c_{2\alpha}\frac{c_{\alpha}}{m_2}\right)\bar{\psi}_{1R}^\zpr\Gamma\nu'_{La}
+ \lambda\,\lambda_a\eta\left(c_{2\alpha}\frac{s_{\alpha}}{m_1}- s_{2\alpha}\frac{c_{\alpha}}{m_2}\right)\bar{\psi}_{2R}^\zpr\Gamma \nu'_{La} + h.c.
\nnmb
\eeq

\noindent\textbf{Dark Higgs Boson ($\Gamma = \varphi/\sqrt{2})$}
\beq
-\lag &\supset& \bar{\psi}_{RI}^{-}\Gamma(\delta_{I2}\lambda_a\delta_{aJ})\psi^{-}_{LJ}
+ \bar{\psi}^0_{Ri}\Gamma(\delta_{i2}\lambda_a\delta_{aJ})\psi^0_{LJ}
\label{eq:hdcoup}\\
&&\nnmb\\
&=&
\bar{\psi}_{RI}^{\mpr}\Gamma(V_{I2}\lambda_aU^{\dagger}_{aJ})\psi^{-}_{LJ}
+ \bar{\psi}^0_{Ri}\Gamma[(B\mathcal{O})_{i2}\lambda_a(\tilde{\mathcal{O}}^{\dagger}A^{\dagger})_{aJ}]\psi^0_{L_J}
\nnmb\\
&&\nnmb\\
&\simeq& \lambda_a\bar{P}_R^{\mpr}\Gamma e^\prime_{La}
+ \frac{\lambda_a^2\eta}{m_P}\bar{P}_R^{\mpr}\Gamma\bar{P}^{\mpr}_L
-\frac{\lambda_a\lambda_b^2\eta^2}{m_P^2}\bar{e}^\prime_{Rb}\Gamma e_{L a}^{\prime}
\nnmb\\
&& +\lambda_a\left(-s_{\alpha}\bar{\psi}^{\zpr}_{R1}+c_{\alpha}\bar{\psi}^{\zpr}_{R2}\right)\Gamma\nu_{La} \nnmb\\
&&+ \lambda_c^2\eta\big(-s_\alpha\bar{\psi}^\zpr_{R 1}+c_\alpha\bar{\psi}^\zpr_{R2}\big)\Gamma\big(-\frac{s_\alpha}{m_1}\psi^\zpr_{L1}+\frac{c_\alpha}{m_2}\psi^\zpr_{L2}\big)
+ h.c.
\nnmb
\eeq


\bibliographystyle{jhep}
\bibliography{BibFile}

\providecommand{\href}[2]{#2}\begingroup\raggedright\begin{thebibliography}{100}

\bibitem{Hewett:1988xc}
J.L.~Hewett and T.G.~Rizzo, \emph{{Low-Energy Phenomenology of Superstring
  Inspired E(6) Models}},
  \href{https://doi.org/10.1016/0370-1573(89)90071-9}{\emph{Phys. Rept.}
  {\bfseries 183} (1989) 193}.

\bibitem{Aldazabal:2000cn}
G.~Aldazabal, S.~Franco, L.E.~Ibanez, R.~Rabadan and A.M.~Uranga,
  \emph{{Intersecting brane worlds}},
  \href{https://doi.org/10.1088/1126-6708/2001/02/047}{\emph{JHEP} {\bfseries
  02} (2001) 047} [\href{https://arxiv.org/abs/hep-ph/0011132}{{\ttfamily
  hep-ph/0011132}}].

\bibitem{Blumenhagen:2006ci}
R.~Blumenhagen, B.~Kors, D.~Lust and S.~Stieberger, \emph{{Four-dimensional
  String Compactifications with D-Branes, Orientifolds and Fluxes}},
  \href{https://doi.org/10.1016/j.physrep.2007.04.003}{\emph{Phys. Rept.}
  {\bfseries 445} (2007) 1}
  [\href{https://arxiv.org/abs/hep-th/0610327}{{\ttfamily hep-th/0610327}}].

\bibitem{Langacker:2008yv}
P.~Langacker, \emph{{The Physics of Heavy $Z^\prime$ Gauge Bosons}},
  \href{https://doi.org/10.1103/RevModPhys.81.1199}{\emph{Rev. Mod. Phys.}
  {\bfseries 81} (2009) 1199}
  [\href{https://arxiv.org/abs/0801.1345}{{\ttfamily 0801.1345}}].

\bibitem{Alexander:2016aln}
J.~Alexander et~al., \emph{{Dark Sectors 2016 Workshop: Community Report}},  8,
  2016 [\href{https://arxiv.org/abs/1608.08632}{{\ttfamily 1608.08632}}].

\bibitem{Boehm:2003hm}
C.~Boehm and P.~Fayet, \emph{{Scalar dark matter candidates}},
  \href{https://doi.org/10.1016/j.nuclphysb.2004.01.015}{\emph{Nucl. Phys. B}
  {\bfseries 683} (2004) 219}
  [\href{https://arxiv.org/abs/hep-ph/0305261}{{\ttfamily hep-ph/0305261}}].

\bibitem{Borodatchenkova:2005ct}
N.~Borodatchenkova, D.~Choudhury and M.~Drees, \emph{{Probing MeV dark matter
  at low-energy e+e- colliders}},
  \href{https://doi.org/10.1103/PhysRevLett.96.141802}{\emph{Phys. Rev. Lett.}
  {\bfseries 96} (2006) 141802}
  [\href{https://arxiv.org/abs/hep-ph/0510147}{{\ttfamily hep-ph/0510147}}].

\bibitem{Pospelov:2007mp}
M.~Pospelov, A.~Ritz and M.B.~Voloshin, \emph{{Secluded WIMP Dark Matter}},
  \href{https://doi.org/10.1016/j.physletb.2008.02.052}{\emph{Phys. Lett. B}
  {\bfseries 662} (2008) 53} [\href{https://arxiv.org/abs/0711.4866}{{\ttfamily
  0711.4866}}].

\bibitem{Arkani-Hamed:2008hhe}
N.~Arkani-Hamed, D.P.~Finkbeiner, T.R.~Slatyer and N.~Weiner, \emph{{A Theory
  of Dark Matter}},
  \href{https://doi.org/10.1103/PhysRevD.79.015014}{\emph{Phys. Rev. D}
  {\bfseries 79} (2009) 015014}
  [\href{https://arxiv.org/abs/0810.0713}{{\ttfamily 0810.0713}}].

\bibitem{Pospelov:2008jd}
M.~Pospelov and A.~Ritz, \emph{{Astrophysical Signatures of Secluded Dark
  Matter}}, \href{https://doi.org/10.1016/j.physletb.2008.12.012}{\emph{Phys.
  Lett. B} {\bfseries 671} (2009) 391}
  [\href{https://arxiv.org/abs/0810.1502}{{\ttfamily 0810.1502}}].

\bibitem{Pospelov:2008zw}
M.~Pospelov, \emph{{Secluded U(1) below the weak scale}},
  \href{https://doi.org/10.1103/PhysRevD.80.095002}{\emph{Phys. Rev. D}
  {\bfseries 80} (2009) 095002}
  [\href{https://arxiv.org/abs/0811.1030}{{\ttfamily 0811.1030}}].

\bibitem{Bjorken:2009mm}
J.D.~Bjorken, R.~Essig, P.~Schuster and N.~Toro, \emph{{New Fixed-Target
  Experiments to Search for Dark Gauge Forces}},
  \href{https://doi.org/10.1103/PhysRevD.80.075018}{\emph{Phys. Rev. D}
  {\bfseries 80} (2009) 075018}
  [\href{https://arxiv.org/abs/0906.0580}{{\ttfamily 0906.0580}}].

\bibitem{Okun:1982xi}
L.B.~Okun, \emph{{LIMITS OF ELECTRODYNAMICS: PARAPHOTONS?}}, {\emph{Sov. Phys.
  JETP} {\bfseries 56} (1982) 502}.

\bibitem{Holdom:1985ag}
B.~Holdom, \emph{{Two U(1)'s and Epsilon Charge Shifts}},
  \href{https://doi.org/10.1016/0370-2693(86)91377-8}{\emph{Phys. Lett. B}
  {\bfseries 166} (1986) 196}.

\bibitem{Faraggi:2000pv}
A.E.~Faraggi and M.~Pospelov, \emph{{Selfinteracting dark matter from the
  hidden heterotic string sector}},
  \href{https://doi.org/10.1016/S0927-6505(01)00121-9}{\emph{Astropart. Phys.}
  {\bfseries 16} (2002) 451}
  [\href{https://arxiv.org/abs/hep-ph/0008223}{{\ttfamily hep-ph/0008223}}].

\bibitem{Juknevich:2009ji}
J.E.~Juknevich, D.~Melnikov and M.J.~Strassler, \emph{{A Pure-Glue Hidden
  Valley I. States and Decays}},
  \href{https://doi.org/10.1088/1126-6708/2009/07/055}{\emph{JHEP} {\bfseries
  07} (2009) 055} [\href{https://arxiv.org/abs/0903.0883}{{\ttfamily
  0903.0883}}].

\bibitem{Juknevich:2009gg}
J.E.~Juknevich, \emph{{Pure-glue hidden valleys through the Higgs portal}},
  \href{https://doi.org/10.1007/JHEP08(2010)121}{\emph{JHEP} {\bfseries 08}
  (2010) 121} [\href{https://arxiv.org/abs/0911.5616}{{\ttfamily 0911.5616}}].

\bibitem{Forestell:2017wov}
L.~Forestell, D.E.~Morrissey and K.~Sigurdson, \emph{{Cosmological Bounds on
  Non-Abelian Dark Forces}},
  \href{https://doi.org/10.1103/PhysRevD.97.075029}{\emph{Phys. Rev. D}
  {\bfseries 97} (2018) 075029}
  [\href{https://arxiv.org/abs/1710.06447}{{\ttfamily 1710.06447}}].

\bibitem{Perl:2009zz}
M.L.~Perl, E.R.~Lee and D.~Loomba, \emph{{Searches for fractionally charged
  particles}},
  \href{https://doi.org/10.1146/annurev-nucl-121908-122035}{\emph{Ann. Rev.
  Nucl. Part. Sci.} {\bfseries 59} (2009) 47}.

\bibitem{DeLuca:2018mzn}
V.~De~Luca, A.~Mitridate, M.~Redi, J.~Smirnov and A.~Strumia, \emph{{Colored
  Dark Matter}}, \href{https://doi.org/10.1103/PhysRevD.97.115024}{\emph{Phys.
  Rev. D} {\bfseries 97} (2018) 115024}
  [\href{https://arxiv.org/abs/1801.01135}{{\ttfamily 1801.01135}}].

\bibitem{Lee:1977ua}
B.W.~Lee and S.~Weinberg, \emph{{Cosmological Lower Bound on Heavy Neutrino
  Masses}}, \href{https://doi.org/10.1103/PhysRevLett.39.165}{\emph{Phys. Rev.
  Lett.} {\bfseries 39} (1977) 165}.

\bibitem{Davoudiasl:2012ig}
H.~Davoudiasl, H.-S.~Lee and W.J.~Marciano, \emph{{Dark Side of Higgs Diphoton
  Decays and Muon g-2}},
  \href{https://doi.org/10.1103/PhysRevD.86.095009}{\emph{Phys. Rev. D}
  {\bfseries 86} (2012) 095009}
  [\href{https://arxiv.org/abs/1208.2973}{{\ttfamily 1208.2973}}].

\bibitem{Lu:2017uur}
Q.~Lu, D.E.~Morrissey and A.M.~Wijangco, \emph{{Higgs Boson Decays to Dark
  Photons through the Vectorized Lepton Portal}},
  \href{https://doi.org/10.1007/JHEP06(2017)138}{\emph{JHEP} {\bfseries 06}
  (2017) 138} [\href{https://arxiv.org/abs/1705.08896}{{\ttfamily
  1705.08896}}].

\bibitem{Kawasaki:2017bqm}
M.~Kawasaki, K.~Kohri, T.~Moroi and Y.~Takaesu, \emph{{Revisiting Big-Bang
  Nucleosynthesis Constraints on Long-Lived Decaying Particles}},
  \href{https://doi.org/10.1103/PhysRevD.97.023502}{\emph{Phys. Rev. D}
  {\bfseries 97} (2018) 023502}
  [\href{https://arxiv.org/abs/1709.01211}{{\ttfamily 1709.01211}}].

\bibitem{Slatyer:2016qyl}
T.R.~Slatyer and C.-L.~Wu, \emph{{General Constraints on Dark Matter Decay from
  the Cosmic Microwave Background}},
  \href{https://doi.org/10.1103/PhysRevD.95.023010}{\emph{Phys. Rev. D}
  {\bfseries 95} (2017) 023010}
  [\href{https://arxiv.org/abs/1610.06933}{{\ttfamily 1610.06933}}].

\bibitem{Cirelli:2005uq}
M.~Cirelli, N.~Fornengo and A.~Strumia, \emph{{Minimal dark matter}},
  \href{https://doi.org/10.1016/j.nuclphysb.2006.07.012}{\emph{Nucl. Phys. B}
  {\bfseries 753} (2006) 178}
  [\href{https://arxiv.org/abs/hep-ph/0512090}{{\ttfamily hep-ph/0512090}}].

\bibitem{Azatov:2011ht}
A.~Azatov, J.~Galloway and M.A.~Luty, \emph{{Superconformal Technicolor}},
  \href{https://doi.org/10.1103/PhysRevLett.108.041802}{\emph{Phys. Rev. Lett.}
  {\bfseries 108} (2012) 041802}
  [\href{https://arxiv.org/abs/1106.3346}{{\ttfamily 1106.3346}}].

\bibitem{Heckman:2011bb}
J.J.~Heckman, P.~Kumar, C.~Vafa and B.~Wecht, \emph{{Electroweak Symmetry
  Breaking in the DSSM}},
  \href{https://doi.org/10.1007/JHEP01(2012)156}{\emph{JHEP} {\bfseries 01}
  (2012) 156} [\href{https://arxiv.org/abs/1108.3849}{{\ttfamily 1108.3849}}].

\bibitem{Graham:2015cka}
P.W.~Graham, D.E.~Kaplan and S.~Rajendran, \emph{{Cosmological Relaxation of
  the Electroweak Scale}},
  \href{https://doi.org/10.1103/PhysRevLett.115.221801}{\emph{Phys. Rev. Lett.}
  {\bfseries 115} (2015) 221801}
  [\href{https://arxiv.org/abs/1504.07551}{{\ttfamily 1504.07551}}].

\bibitem{Beauchesne:2017ukw}
H.~Beauchesne, E.~Bertuzzo and G.~Grilli~di Cortona, \emph{{Constraints on the
  relaxion mechanism with strongly interacting vector-fermions}},
  \href{https://doi.org/10.1007/JHEP08(2017)093}{\emph{JHEP} {\bfseries 08}
  (2017) 093} [\href{https://arxiv.org/abs/1705.06325}{{\ttfamily
  1705.06325}}].

\bibitem{Wojcik:2020wgm}
G.N.~Wojcik and T.G.~Rizzo, \emph{{SU(4) flavorful portal matter}},
  \href{https://doi.org/10.1103/PhysRevD.105.015032}{\emph{Phys. Rev. D}
  {\bfseries 105} (2022) 015032}
  [\href{https://arxiv.org/abs/2012.05406}{{\ttfamily 2012.05406}}].

\bibitem{Rizzo:2022lpm}
T.G.~Rizzo, \emph{{Towards a UV-Model of Kinetic Mixing and Portal Matter:
  Exploring Unification in an $SU(N)$ Group}},
  \href{https://arxiv.org/abs/2209.00688}{{\ttfamily 2209.00688}}.

\bibitem{Cline:2016nab}
J.M.~Cline, W.~Huang and G.D.~Moore, \emph{{Challenges for models with
  composite states}},
  \href{https://doi.org/10.1103/PhysRevD.94.055029}{\emph{Phys. Rev. D}
  {\bfseries 94} (2016) 055029}
  [\href{https://arxiv.org/abs/1607.07865}{{\ttfamily 1607.07865}}].

\bibitem{Carone:2018eka}
C.D.~Carone, S.~Chaurasia and T.V.B.~Claringbold, \emph{{Dark sector portal
  with vectorlike leptons and flavor sequestering}},
  \href{https://doi.org/10.1103/PhysRevD.99.015009}{\emph{Phys. Rev. D}
  {\bfseries 99} (2019) 015009}
  [\href{https://arxiv.org/abs/1807.05288}{{\ttfamily 1807.05288}}].

\bibitem{Rizzo:2018vlb}
T.G.~Rizzo, \emph{{Kinetic Mixing and Portal Matter Phenomenology}},
  \href{https://doi.org/10.1103/PhysRevD.99.115024}{\emph{Phys. Rev. D}
  {\bfseries 99} (2019) 115024}
  [\href{https://arxiv.org/abs/1810.07531}{{\ttfamily 1810.07531}}].

\bibitem{Kim:2019oyh}
J.H.~Kim, S.D.~Lane, H.-S.~Lee, I.M.~Lewis and M.~Sullivan, \emph{{Searching
  for Dark Photons with Maverick Top Partners}},
  \href{https://doi.org/10.1103/PhysRevD.101.035041}{\emph{Phys. Rev. D}
  {\bfseries 101} (2020) 035041}
  [\href{https://arxiv.org/abs/1904.05893}{{\ttfamily 1904.05893}}].

\bibitem{Lamprea:2019qet}
J.M.~Lamprea, E.~Peinado, S.~Smolenski and J.~Wudka, \emph{{Self-interacting
  neutrino portal dark matter}},
  \href{https://doi.org/10.1103/PhysRevD.103.015017}{\emph{Phys. Rev. D}
  {\bfseries 103} (2021) 015017}
  [\href{https://arxiv.org/abs/1906.02340}{{\ttfamily 1906.02340}}].

\bibitem{Rueter:2019wdf}
T.D.~Rueter and T.G.~Rizzo, \emph{{Towards A UV-Model of Kinetic Mixing and
  Portal Matter}},
  \href{https://doi.org/10.1103/PhysRevD.101.015014}{\emph{Phys. Rev. D}
  {\bfseries 101} (2020) 015014}
  [\href{https://arxiv.org/abs/1909.09160}{{\ttfamily 1909.09160}}].

\bibitem{Coy:2020wxp}
R.~Coy and T.~Hambye, \emph{{Neutrino lines from DM decay induced by high-scale
  seesaw interactions}},
  \href{https://doi.org/10.1007/JHEP05(2021)101}{\emph{JHEP} {\bfseries 05}
  (2021) 101} [\href{https://arxiv.org/abs/2012.05276}{{\ttfamily
  2012.05276}}].

\bibitem{Banta:2021dek}
I.~Banta, T.~Cohen, N.~Craig, X.~Lu and D.~Sutherland, \emph{{Non-decoupling
  new particles}}, \href{https://doi.org/10.1007/JHEP02(2022)029}{\emph{JHEP}
  {\bfseries 02} (2022) 029}
  [\href{https://arxiv.org/abs/2110.02967}{{\ttfamily 2110.02967}}].

\bibitem{Stueckelberg:1957zz}
E.C.G.~Stueckelberg, \emph{{Theory of the radiation of photons of small
  arbitrary mass}}, {\emph{Helv. Phys. Acta} {\bfseries 30} (1957) 209}.

\bibitem{Kors:2004dx}
B.~Kors and P.~Nath, \emph{{A Stueckelberg extension of the standard model}},
  \href{https://doi.org/10.1016/j.physletb.2004.02.051}{\emph{Phys. Lett. B}
  {\bfseries 586} (2004) 366}
  [\href{https://arxiv.org/abs/hep-ph/0402047}{{\ttfamily hep-ph/0402047}}].

\bibitem{DiFranzo:2015nli}
A.~DiFranzo, P.J.~Fox and T.M.P.~Tait, \emph{{Vector Dark Matter through a
  Radiative Higgs Portal}},
  \href{https://doi.org/10.1007/JHEP04(2016)135}{\emph{JHEP} {\bfseries 04}
  (2016) 135} [\href{https://arxiv.org/abs/1512.06853}{{\ttfamily
  1512.06853}}].

\bibitem{DiFranzo:2016uzc}
A.~DiFranzo and G.~Mohlabeng, \emph{{Multi-component Dark Matter through a
  Radiative Higgs Portal}},
  \href{https://doi.org/10.1007/JHEP01(2017)080}{\emph{JHEP} {\bfseries 01}
  (2017) 080} [\href{https://arxiv.org/abs/1610.07606}{{\ttfamily
  1610.07606}}].

\bibitem{Gherghetta:2019coi}
T.~Gherghetta, J.~Kersten, K.~Olive and M.~Pospelov, \emph{{Evaluating the
  price of tiny kinetic mixing}},
  \href{https://doi.org/10.1103/PhysRevD.100.095001}{\emph{Phys. Rev. D}
  {\bfseries 100} (2019) 095001}
  [\href{https://arxiv.org/abs/1909.00696}{{\ttfamily 1909.00696}}].

\bibitem{Shifman:1979eb}
M.A.~Shifman, A.I.~Vainshtein, M.B.~Voloshin and V.I.~Zakharov,
  \emph{{Low-Energy Theorems for Higgs Boson Couplings to Photons}},
  {\emph{Sov. J. Nucl. Phys.} {\bfseries 30} (1979) 711}.

\bibitem{Agrawal:2021dbo}
P.~Agrawal et~al., \emph{{Feebly-interacting particles: FIPs 2020 workshop
  report}}, \href{https://doi.org/10.1140/epjc/s10052-021-09703-7}{\emph{Eur.
  Phys. J. C} {\bfseries 81} (2021) 1015}
  [\href{https://arxiv.org/abs/2102.12143}{{\ttfamily 2102.12143}}].

\bibitem{LHCb:2019vmc}
{\scshape LHCb} collaboration, \emph{{Search for $A'\to\mu^+\mu^-$ Decays}},
  \href{https://doi.org/10.1103/PhysRevLett.124.041801}{\emph{Phys. Rev. Lett.}
  {\bfseries 124} (2020) 041801}
  [\href{https://arxiv.org/abs/1910.06926}{{\ttfamily 1910.06926}}].

\bibitem{CMS:2019buh}
{\scshape CMS} collaboration, \emph{{Search for a Narrow Resonance Lighter than
  200 GeV Decaying to a Pair of Muons in Proton-Proton Collisions at $\sqrt{s}
  =$ TeV}}, \href{https://doi.org/10.1103/PhysRevLett.124.131802}{\emph{Phys.
  Rev. Lett.} {\bfseries 124} (2020) 131802}
  [\href{https://arxiv.org/abs/1912.04776}{{\ttfamily 1912.04776}}].

\bibitem{Peskin:1990zt}
M.E.~Peskin and T.~Takeuchi, \emph{{A New constraint on a strongly interacting
  Higgs sector}}, \href{https://doi.org/10.1103/PhysRevLett.65.964}{\emph{Phys.
  Rev. Lett.} {\bfseries 65} (1990) 964}.

\bibitem{Peskin:1991sw}
M.E.~Peskin and T.~Takeuchi, \emph{{Estimation of oblique electroweak
  corrections}}, \href{https://doi.org/10.1103/PhysRevD.46.381}{\emph{Phys.
  Rev. D} {\bfseries 46} (1992) 381}.

\bibitem{ParticleDataGroup:2020ssz}
{\scshape Particle Data Group} collaboration, \emph{{Review of Particle
  Physics}}, \href{https://doi.org/10.1093/ptep/ptaa104}{\emph{PTEP} {\bfseries
  2020} (2020) 083C01}.

\bibitem{CDF:2022hxs}
{\scshape CDF} collaboration, \emph{{High-precision measurement of the W boson
  mass with the CDF II detector}},
  \href{https://doi.org/10.1126/science.abk1781}{\emph{Science} {\bfseries 376}
  (2022) 170}.

\bibitem{ATLAS:2022yvh}
{\scshape ATLAS} collaboration, \emph{{Search for invisible Higgs-boson decays
  in events with vector-boson fusion signatures using 139 fb$^{-1}$ of
  proton-proton data recorded by the ATLAS experiment}},
  \href{https://doi.org/10.1007/JHEP08(2022)104}{\emph{JHEP} {\bfseries 08}
  (2022) 104} [\href{https://arxiv.org/abs/2202.07953}{{\ttfamily
  2202.07953}}].

\bibitem{ATLAS:2021ldb}
{\scshape ATLAS} collaboration, \emph{{Search for Higgs bosons decaying into
  new spin-0 or spin-1 particles in four-lepton final states with the ATLAS
  detector with 139 fb$^{-1}$ of $pp$ collision data at $\sqrt{s}=13$ TeV}},
  \href{https://doi.org/10.1007/JHEP03(2022)041}{\emph{JHEP} {\bfseries 03}
  (2022) 041} [\href{https://arxiv.org/abs/2110.13673}{{\ttfamily
  2110.13673}}].

\bibitem{CMS:2021pcy}
{\scshape CMS} collaboration, \emph{{Search for low-mass dilepton resonances in
  Higgs boson decays to four-lepton final states in proton\textendash{}proton
  collisions at $\sqrt{s}=13\,\text {TeV} $}},
  \href{https://doi.org/10.1140/epjc/s10052-022-10127-0}{\emph{Eur. Phys. J. C}
  {\bfseries 82} (2022) 290}
  [\href{https://arxiv.org/abs/2111.01299}{{\ttfamily 2111.01299}}].

\bibitem{Curtin:2014cca}
D.~Curtin, R.~Essig, S.~Gori and J.~Shelton, \emph{{Illuminating Dark Photons
  with High-Energy Colliders}},
  \href{https://doi.org/10.1007/JHEP02(2015)157}{\emph{JHEP} {\bfseries 02}
  (2015) 157} [\href{https://arxiv.org/abs/1412.0018}{{\ttfamily 1412.0018}}].

\bibitem{Canepa:2020ntc}
A.~Canepa, T.~Han and X.~Wang, \emph{{The Search for Electroweakinos}},
  \href{https://doi.org/10.1146/annurev-nucl-031020-121031}{\emph{Ann. Rev.
  Nucl. Part. Sci.} {\bfseries 70} (2020) 425}
  [\href{https://arxiv.org/abs/2003.05450}{{\ttfamily 2003.05450}}].

\bibitem{Cohen:2011ec}
T.~Cohen, J.~Kearney, A.~Pierce and D.~Tucker-Smith, \emph{{Singlet-Doublet
  Dark Matter}}, \href{https://doi.org/10.1103/PhysRevD.85.075003}{\emph{Phys.
  Rev. D} {\bfseries 85} (2012) 075003}
  [\href{https://arxiv.org/abs/1109.2604}{{\ttfamily 1109.2604}}].

\bibitem{Martin:2014qra}
T.A.W.~Martin and D.~Morrissey, \emph{{Electroweakino constraints from LHC
  data}}, \href{https://doi.org/10.1007/JHEP12(2014)168}{\emph{JHEP} {\bfseries
  12} (2014) 168} [\href{https://arxiv.org/abs/1409.6322}{{\ttfamily
  1409.6322}}].

\bibitem{Liu:2020muv}
J.~Liu, N.~McGinnis, C.E.M.~Wagner and X.-P.~Wang, \emph{{Searching for the
  Higgsino-Bino Sector at the LHC}},
  \href{https://doi.org/10.1007/JHEP09(2020)073}{\emph{JHEP} {\bfseries 09}
  (2020) 073} [\href{https://arxiv.org/abs/2006.07389}{{\ttfamily
  2006.07389}}].

\bibitem{ATLAS:2021yqv}
{\scshape ATLAS} collaboration, \emph{{Search for charginos and neutralinos in
  final states with two boosted hadronically decaying bosons and missing
  transverse momentum in $pp$ collisions at $\sqrt {s}$ = 13\,\,TeV with the
  ATLAS detector}},
  \href{https://doi.org/10.1103/PhysRevD.104.112010}{\emph{Phys. Rev. D}
  {\bfseries 104} (2021) 112010}
  [\href{https://arxiv.org/abs/2108.07586}{{\ttfamily 2108.07586}}].

\bibitem{Alwall:2014hca}
J.~Alwall, R.~Frederix, S.~Frixione, V.~Hirschi, F.~Maltoni, O.~Mattelaer
  et~al., \emph{{The automated computation of tree-level and next-to-leading
  order differential cross sections, and their matching to parton shower
  simulations}}, \href{https://doi.org/10.1007/JHEP07(2014)079}{\emph{JHEP}
  {\bfseries 07} (2014) 079} [\href{https://arxiv.org/abs/1405.0301}{{\ttfamily
  1405.0301}}].

\bibitem{Alloul:2013bka}
A.~Alloul, N.D.~Christensen, C.~Degrande, C.~Duhr and B.~Fuks, \emph{{FeynRules
  2.0 - A complete toolbox for tree-level phenomenology}},
  \href{https://doi.org/10.1016/j.cpc.2014.04.012}{\emph{Comput. Phys. Commun.}
  {\bfseries 185} (2014) 2250}
  [\href{https://arxiv.org/abs/1310.1921}{{\ttfamily 1310.1921}}].

\bibitem{hepdata.104458}
{ATLAS Collaboration}, ``{Search for charginos and neutralinos in final states
  with two boosted hadronically decaying bosons and missing transverse momentum
  in $pp$ collisions at $\sqrt{s}=13$\textasciitilde{}TeV with the ATLAS
  detector}.'' {HEPData (collection)}, 2021.

\bibitem{Cowan:2010js}
G.~Cowan, K.~Cranmer, E.~Gross and O.~Vitells, \emph{{Asymptotic formulae for
  likelihood-based tests of new physics}},
  \href{https://doi.org/10.1140/epjc/s10052-011-1554-0}{\emph{Eur. Phys. J. C}
  {\bfseries 71} (2011) 1554}
  [\href{https://arxiv.org/abs/1007.1727}{{\ttfamily 1007.1727}}].

\bibitem{Christensen:2008py}
N.D.~Christensen and C.~Duhr, \emph{{FeynRules - Feynman rules made easy}},
  \href{https://doi.org/10.1016/j.cpc.2009.02.018}{\emph{Comput. Phys. Commun.}
  {\bfseries 180} (2009) 1614}
  [\href{https://arxiv.org/abs/0806.4194}{{\ttfamily 0806.4194}}].

\bibitem{Degrande:2011ua}
C.~Degrande, C.~Duhr, B.~Fuks, D.~Grellscheid, O.~Mattelaer and T.~Reiter,
  \emph{{UFO - The Universal FeynRules Output}},
  \href{https://doi.org/10.1016/j.cpc.2012.01.022}{\emph{Comput. Phys. Commun.}
  {\bfseries 183} (2012) 1201}
  [\href{https://arxiv.org/abs/1108.2040}{{\ttfamily 1108.2040}}].

\bibitem{Backovic:2013dpa}
M.~Backovic, K.~Kong and M.~McCaskey, \emph{{MadDM v.1.0: Computation of Dark
  Matter Relic Abundance Using MadGraph5}},
  \href{https://doi.org/10.1016/j.dark.2014.04.001}{\emph{Physics of the Dark
  Universe} {\bfseries 5-6} (2014) 18}
  [\href{https://arxiv.org/abs/1308.4955}{{\ttfamily 1308.4955}}].

\bibitem{Backovic:2015cra}
M.~Backovi\'c, A.~Martini, O.~Mattelaer, K.~Kong and G.~Mohlabeng,
  \emph{{Direct Detection of Dark Matter with MadDM v.2.0}},
  \href{https://doi.org/10.1016/j.dark.2015.09.001}{\emph{Phys. Dark Univ.}
  {\bfseries 9-10} (2015) 37}
  [\href{https://arxiv.org/abs/1505.04190}{{\ttfamily 1505.04190}}].

\bibitem{Ambrogi:2018jqj}
F.~Ambrogi, C.~Arina, M.~Backovic, J.~Heisig, F.~Maltoni, L.~Mantani et~al.,
  \emph{{MadDM v.3.0: a Comprehensive Tool for Dark Matter Studies}},
  \href{https://doi.org/10.1016/j.dark.2018.11.009}{\emph{Phys. Dark Univ.}
  {\bfseries 24} (2019) 100249}
  [\href{https://arxiv.org/abs/1804.00044}{{\ttfamily 1804.00044}}].

\bibitem{Cassel:2009wt}
S.~Cassel, \emph{{Sommerfeld factor for arbitrary partial wave processes}},
  \href{https://doi.org/10.1088/0954-3899/37/10/105009}{\emph{J. Phys. G}
  {\bfseries 37} (2010) 105009}
  [\href{https://arxiv.org/abs/0903.5307}{{\ttfamily 0903.5307}}].

\bibitem{Slatyer:2009vg}
T.R.~Slatyer, \emph{{The Sommerfeld enhancement for dark matter with an excited
  state}}, \href{https://doi.org/10.1088/1475-7516/2010/02/028}{\emph{JCAP}
  {\bfseries 02} (2010) 028} [\href{https://arxiv.org/abs/0910.5713}{{\ttfamily
  0910.5713}}].

\bibitem{Mizuta:1992qp}
S.~Mizuta and M.~Yamaguchi, \emph{{Coannihilation effects and relic abundance
  of Higgsino dominant LSP(s)}},
  \href{https://doi.org/10.1016/0370-2693(93)91717-2}{\emph{Phys. Lett. B}
  {\bfseries 298} (1993) 120}
  [\href{https://arxiv.org/abs/hep-ph/9208251}{{\ttfamily hep-ph/9208251}}].

\bibitem{Cirelli:2007xd}
M.~Cirelli, A.~Strumia and M.~Tamburini, \emph{{Cosmology and Astrophysics of
  Minimal Dark Matter}},
  \href{https://doi.org/10.1016/j.nuclphysb.2007.07.023}{\emph{Nucl. Phys. B}
  {\bfseries 787} (2007) 152}
  [\href{https://arxiv.org/abs/0706.4071}{{\ttfamily 0706.4071}}].

\bibitem{Dessert:2022evk}
C.~Dessert, J.W.~Foster, Y.~Park, B.R.~Safdi and W.L.~Xu, \emph{{Higgsino Dark
  Matter Confronts 14 years of Fermi Gamma Ray Data}},
  \href{https://arxiv.org/abs/2207.10090}{{\ttfamily 2207.10090}}.

\bibitem{Jungman:1995df}
G.~Jungman, M.~Kamionkowski and K.~Griest, \emph{{Supersymmetric dark matter}},
  \href{https://doi.org/10.1016/0370-1573(95)00058-5}{\emph{Phys. Rept.}
  {\bfseries 267} (1996) 195}
  [\href{https://arxiv.org/abs/hep-ph/9506380}{{\ttfamily hep-ph/9506380}}].

\bibitem{DelNobile:2021wmp}
E.~Del~Nobile, \emph{{The Theory of Direct Dark Matter Detection: A Guide to
  Computations}},  \href{https://arxiv.org/abs/2104.12785}{{\ttfamily
  2104.12785}}.

\bibitem{LUX-ZEPLIN:2022qhg}
{\scshape LZ} collaboration, \emph{{First Dark Matter Search Results from the
  LUX-ZEPLIN (LZ) Experiment}},
  \href{https://arxiv.org/abs/2207.03764}{{\ttfamily 2207.03764}}.

\bibitem{Tucker-Smith:2001myb}
D.~Tucker-Smith and N.~Weiner, \emph{{Inelastic dark matter}},
  \href{https://doi.org/10.1103/PhysRevD.64.043502}{\emph{Phys. Rev. D}
  {\bfseries 64} (2001) 043502}
  [\href{https://arxiv.org/abs/hep-ph/0101138}{{\ttfamily hep-ph/0101138}}].

\bibitem{Bramante:2016rdh}
J.~Bramante, P.J.~Fox, G.D.~Kribs and A.~Martin, \emph{{Inelastic frontier:
  Discovering dark matter at high recoil energy}},
  \href{https://doi.org/10.1103/PhysRevD.94.115026}{\emph{Phys. Rev. D}
  {\bfseries 94} (2016) 115026}
  [\href{https://arxiv.org/abs/1608.02662}{{\ttfamily 1608.02662}}].

\bibitem{PandaX:2022djq}
{\scshape PandaX} collaboration, \emph{{A search for two-component Majorana
  dark matter in a simplified model using the full exposure data of PandaX-II
  experiment}},
  \href{https://doi.org/10.1016/j.physletb.2022.137254}{\emph{Phys. Lett. B}
  {\bfseries 832} (2022) 137254}
  [\href{https://arxiv.org/abs/2205.08066}{{\ttfamily 2205.08066}}].

\bibitem{Finkbeiner:2009mi}
D.P.~Finkbeiner, T.R.~Slatyer, N.~Weiner and I.~Yavin, \emph{{PAMELA, DAMA,
  INTEGRAL and Signatures of Metastable Excited WIMPs}},
  \href{https://doi.org/10.1088/1475-7516/2009/09/037}{\emph{JCAP} {\bfseries
  09} (2009) 037} [\href{https://arxiv.org/abs/0903.1037}{{\ttfamily
  0903.1037}}].

\bibitem{CarrilloGonzalez:2021lxm}
M.~Carrillo~Gonz\'alez and N.~Toro, \emph{{Cosmology and signals of light
  pseudo-Dirac dark matter}},
  \href{https://doi.org/10.1007/JHEP04(2022)060}{\emph{JHEP} {\bfseries 04}
  (2022) 060} [\href{https://arxiv.org/abs/2108.13422}{{\ttfamily
  2108.13422}}].

\bibitem{Song:2021yar}
N.~Song, S.~Nagorny and A.C.~Vincent, \emph{{Pushing the frontier of WIMPy
  inelastic dark matter: Journey to the end of the periodic table}},
  \href{https://doi.org/10.1103/PhysRevD.104.103032}{\emph{Phys. Rev. D}
  {\bfseries 104} (2021) 103032}
  [\href{https://arxiv.org/abs/2104.09517}{{\ttfamily 2104.09517}}].

\bibitem{Hess:2021cdp}
{\scshape Hess, HAWC, VERITAS, MAGIC, H.E.S.S., Fermi-LAT} collaboration,
  \emph{{Combined dark matter searches towards dwarf spheroidal galaxies with
  Fermi-LAT, HAWC, H.E.S.S., MAGIC, and VERITAS}},
  \href{https://doi.org/10.22323/1.395.0528}{\emph{PoS} {\bfseries ICRC2021}
  (2021) 528} [\href{https://arxiv.org/abs/2108.13646}{{\ttfamily
  2108.13646}}].

\bibitem{Fermi-LAT:2017opo}
{\scshape Fermi-LAT} collaboration, \emph{{The Fermi Galactic Center GeV Excess
  and Implications for Dark Matter}},
  \href{https://doi.org/10.3847/1538-4357/aa6cab}{\emph{Astrophys. J.}
  {\bfseries 840} (2017) 43}
  [\href{https://arxiv.org/abs/1704.03910}{{\ttfamily 1704.03910}}].

\bibitem{HESS:2022ygk}
{\scshape H.E.S.S.} collaboration, \emph{{Search for dark matter annihilation
  signals in the H.E.S.S. Inner Galaxy Survey}},
  \href{https://arxiv.org/abs/2207.10471}{{\ttfamily 2207.10471}}.

\bibitem{Chen:2003gz}
X.-L.~Chen and M.~Kamionkowski, \emph{{Particle decays during the cosmic dark
  ages}}, \href{https://doi.org/10.1103/PhysRevD.70.043502}{\emph{Phys. Rev. D}
  {\bfseries 70} (2004) 043502}
  [\href{https://arxiv.org/abs/astro-ph/0310473}{{\ttfamily
  astro-ph/0310473}}].

\bibitem{Padmanabhan:2005es}
N.~Padmanabhan and D.P.~Finkbeiner, \emph{{Detecting dark matter annihilation
  with CMB polarization: Signatures and experimental prospects}},
  \href{https://doi.org/10.1103/PhysRevD.72.023508}{\emph{Phys. Rev. D}
  {\bfseries 72} (2005) 023508}
  [\href{https://arxiv.org/abs/astro-ph/0503486}{{\ttfamily
  astro-ph/0503486}}].

\bibitem{Slatyer:2009yq}
T.R.~Slatyer, N.~Padmanabhan and D.P.~Finkbeiner, \emph{{CMB Constraints on
  WIMP Annihilation: Energy Absorption During the Recombination Epoch}},
  \href{https://doi.org/10.1103/PhysRevD.80.043526}{\emph{Phys. Rev. D}
  {\bfseries 80} (2009) 043526}
  [\href{https://arxiv.org/abs/0906.1197}{{\ttfamily 0906.1197}}].

\bibitem{Planck:2018vyg}
{\scshape Planck} collaboration, \emph{{Planck 2018 results. VI. Cosmological
  parameters}},
  \href{https://doi.org/10.1051/0004-6361/201833910}{\emph{Astron. Astrophys.}
  {\bfseries 641} (2020) A6}
  [\href{https://arxiv.org/abs/1807.06209}{{\ttfamily 1807.06209}}].

\bibitem{Cline:2013fm}
J.M.~Cline and P.~Scott, \emph{{Dark Matter CMB Constraints and Likelihoods for
  Poor Particle Physicists}},
  \href{https://doi.org/10.1088/1475-7516/2013/03/044}{\emph{JCAP} {\bfseries
  03} (2013) 044} [\href{https://arxiv.org/abs/1301.5908}{{\ttfamily
  1301.5908}}].

\bibitem{Slatyer:2015jla}
T.R.~Slatyer, \emph{{Indirect dark matter signatures in the cosmic dark ages.
  I. Generalizing the bound on s-wave dark matter annihilation from Planck
  results}}, \href{https://doi.org/10.1103/PhysRevD.93.023527}{\emph{Phys. Rev.
  D} {\bfseries 93} (2016) 023527}
  [\href{https://arxiv.org/abs/1506.03811}{{\ttfamily 1506.03811}}].

\bibitem{Batell:2009yf}
B.~Batell, M.~Pospelov and A.~Ritz, \emph{{Probing a Secluded U(1) at
  B-factories}}, \href{https://doi.org/10.1103/PhysRevD.79.115008}{\emph{Phys.
  Rev. D} {\bfseries 79} (2009) 115008}
  [\href{https://arxiv.org/abs/0903.0363}{{\ttfamily 0903.0363}}].

\bibitem{Chan:2011aa}
Y.F.~Chan, M.~Low, D.E.~Morrissey and A.P.~Spray, \emph{{LHC Signatures of a
  Minimal Supersymmetric Hidden Valley}},
  \href{https://doi.org/10.1007/JHEP05(2012)155}{\emph{JHEP} {\bfseries 05}
  (2012) 155} [\href{https://arxiv.org/abs/1112.2705}{{\ttfamily 1112.2705}}].

\bibitem{Morrissey:2014yma}
D.E.~Morrissey and A.P.~Spray, \emph{{New Limits on Light Hidden Sectors from
  Fixed-Target Experiments}},
  \href{https://doi.org/10.1007/JHEP06(2014)083}{\emph{JHEP} {\bfseries 06}
  (2014) 083} [\href{https://arxiv.org/abs/1402.4817}{{\ttfamily 1402.4817}}].

\bibitem{Berger:2016vxi}
J.~Berger, K.~Jedamzik and D.G.E.~Walker, \emph{{Cosmological Constraints on
  Decoupled Dark Photons and Dark Higgs}},
  \href{https://doi.org/10.1088/1475-7516/2016/11/032}{\emph{JCAP} {\bfseries
  11} (2016) 032} [\href{https://arxiv.org/abs/1605.07195}{{\ttfamily
  1605.07195}}].

\bibitem{Izaguirre:2015zva}
E.~Izaguirre, G.~Krnjaic and B.~Shuve, \emph{{Discovering Inelastic
  Thermal-Relic Dark Matter at Colliders}},
  \href{https://doi.org/10.1103/PhysRevD.93.063523}{\emph{Phys. Rev. D}
  {\bfseries 93} (2016) 063523}
  [\href{https://arxiv.org/abs/1508.03050}{{\ttfamily 1508.03050}}].

\bibitem{Curtin:2018mvb}
D.~Curtin et~al., \emph{{Long-Lived Particles at the Energy Frontier: The
  MATHUSLA Physics Case}},
  \href{https://doi.org/10.1088/1361-6633/ab28d6}{\emph{Rept. Prog. Phys.}
  {\bfseries 82} (2019) 116201}
  [\href{https://arxiv.org/abs/1806.07396}{{\ttfamily 1806.07396}}].

\bibitem{Berlin:2018jbm}
A.~Berlin and F.~Kling, \emph{{Inelastic Dark Matter at the LHC Lifetime
  Frontier: ATLAS, CMS, LHCb, CODEX-b, FASER, and MATHUSLA}},
  \href{https://doi.org/10.1103/PhysRevD.99.015021}{\emph{Phys. Rev. D}
  {\bfseries 99} (2019) 015021}
  [\href{https://arxiv.org/abs/1810.01879}{{\ttfamily 1810.01879}}].

\bibitem{FASER:2018eoc}
{\scshape FASER} collaboration, \emph{{FASER\textquoteright{}s physics reach
  for long-lived particles}},
  \href{https://doi.org/10.1103/PhysRevD.99.095011}{\emph{Phys. Rev. D}
  {\bfseries 99} (2019) 095011}
  [\href{https://arxiv.org/abs/1811.12522}{{\ttfamily 1811.12522}}].

\bibitem{Cornwall:1974km}
J.M.~Cornwall, D.N.~Levin and G.~Tiktopoulos, \emph{{Derivation of Gauge
  Invariance from High-Energy Unitarity Bounds on the s Matrix}},
  \href{https://doi.org/10.1103/PhysRevD.10.1145}{\emph{Phys. Rev. D}
  {\bfseries 10} (1974) 1145}.

\bibitem{Vayonakis:1976vz}
C.E.~Vayonakis, \emph{{Born Helicity Amplitudes and Cross-Sections in
  Nonabelian Gauge Theories}},
  \href{https://doi.org/10.1007/BF02746538}{\emph{Lett. Nuovo Cim.} {\bfseries
  17} (1976) 383}.

\bibitem{Chanowitz:1985hj}
M.S.~Chanowitz and M.K.~Gaillard, \emph{{The TeV Physics of Strongly
  Interacting W's and Z's}},
  \href{https://doi.org/10.1016/0550-3213(85)90580-2}{\emph{Nucl. Phys. B}
  {\bfseries 261} (1985) 379}.

\bibitem{Kawasaki:2004qu}
M.~Kawasaki, K.~Kohri and T.~Moroi, \emph{{Big-Bang nucleosynthesis and
  hadronic decay of long-lived massive particles}},
  \href{https://doi.org/10.1103/PhysRevD.71.083502}{\emph{Phys. Rev. D}
  {\bfseries 71} (2005) 083502}
  [\href{https://arxiv.org/abs/astro-ph/0408426}{{\ttfamily
  astro-ph/0408426}}].

\bibitem{Jedamzik:2006xz}
K.~Jedamzik, \emph{{Big bang nucleosynthesis constraints on hadronically and
  electromagnetically decaying relic neutral particles}},
  \href{https://doi.org/10.1103/PhysRevD.74.103509}{\emph{Phys. Rev. D}
  {\bfseries 74} (2006) 103509}
  [\href{https://arxiv.org/abs/hep-ph/0604251}{{\ttfamily hep-ph/0604251}}].

\bibitem{Hisano:1995cp}
J.~Hisano, T.~Moroi, K.~Tobe and M.~Yamaguchi, \emph{{Lepton flavor violation
  via right-handed neutrino Yukawa couplings in supersymmetric standard
  model}}, \href{https://doi.org/10.1103/PhysRevD.53.2442}{\emph{Phys. Rev. D}
  {\bfseries 53} (1996) 2442}
  [\href{https://arxiv.org/abs/hep-ph/9510309}{{\ttfamily hep-ph/9510309}}].

\bibitem{Calibbi:2017uvl}
L.~Calibbi and G.~Signorelli, \emph{{Charged Lepton Flavour Violation: An
  Experimental and Theoretical Introduction}},
  \href{https://doi.org/10.1393/ncr/i2018-10144-0}{\emph{Riv. Nuovo Cim.}
  {\bfseries 41} (2018) 71} [\href{https://arxiv.org/abs/1709.00294}{{\ttfamily
  1709.00294}}].

\bibitem{Calibbi:2018rzv}
L.~Calibbi, R.~Ziegler and J.~Zupan, \emph{{Minimal models for dark matter and
  the muon g\ensuremath{-}2 anomaly}},
  \href{https://doi.org/10.1007/JHEP07(2018)046}{\emph{JHEP} {\bfseries 07}
  (2018) 046} [\href{https://arxiv.org/abs/1804.00009}{{\ttfamily
  1804.00009}}].

\bibitem{Crivellin:2020ebi}
A.~Crivellin, F.~Kirk, C.A.~Manzari and M.~Montull, \emph{{Global Electroweak
  Fit and Vector-Like Leptons in Light of the Cabibbo Angle Anomaly}},
  \href{https://doi.org/10.1007/JHEP12(2020)166}{\emph{JHEP} {\bfseries 12}
  (2020) 166} [\href{https://arxiv.org/abs/2008.01113}{{\ttfamily
  2008.01113}}].

\bibitem{Toma:2013zsa}
T.~Toma and A.~Vicente, \emph{{Lepton Flavor Violation in the Scotogenic
  Model}}, \href{https://doi.org/10.1007/JHEP01(2014)160}{\emph{JHEP}
  {\bfseries 01} (2014) 160} [\href{https://arxiv.org/abs/1312.2840}{{\ttfamily
  1312.2840}}].

\bibitem{Muong-2:2006rrc}
{\scshape Muon g-2} collaboration, \emph{{Final Report of the Muon E821
  Anomalous Magnetic Moment Measurement at BNL}},
  \href{https://doi.org/10.1103/PhysRevD.73.072003}{\emph{Phys. Rev. D}
  {\bfseries 73} (2006) 072003}
  [\href{https://arxiv.org/abs/hep-ex/0602035}{{\ttfamily hep-ex/0602035}}].

\bibitem{Muong-2:2021ojo}
{\scshape Muon g-2} collaboration, \emph{{Measurement of the Positive Muon
  Anomalous Magnetic Moment to 0.46 ppm}},
  \href{https://doi.org/10.1103/PhysRevLett.126.141801}{\emph{Phys. Rev. Lett.}
  {\bfseries 126} (2021) 141801}
  [\href{https://arxiv.org/abs/2104.03281}{{\ttfamily 2104.03281}}].

\bibitem{MEG:2016leq}
{\scshape MEG} collaboration, \emph{{Search for the lepton flavour violating
  decay $\mu ^+ \rightarrow \mathrm {e}^+ \gamma $ with the full dataset of the
  MEG experiment}},
  \href{https://doi.org/10.1140/epjc/s10052-016-4271-x}{\emph{Eur. Phys. J. C}
  {\bfseries 76} (2016) 434}
  [\href{https://arxiv.org/abs/1605.05081}{{\ttfamily 1605.05081}}].

\bibitem{Strassler:2006im}
M.J.~Strassler and K.M.~Zurek, \emph{{Echoes of a hidden valley at hadron
  colliders}},
  \href{https://doi.org/10.1016/j.physletb.2007.06.055}{\emph{Phys. Lett. B}
  {\bfseries 651} (2007) 374}
  [\href{https://arxiv.org/abs/hep-ph/0604261}{{\ttfamily hep-ph/0604261}}].

\bibitem{Strassler:2006qa}
M.J.~Strassler, \emph{{Possible effects of a hidden valley on supersymmetric
  phenomenology}},  \href{https://arxiv.org/abs/hep-ph/0607160}{{\ttfamily
  hep-ph/0607160}}.

\bibitem{Han:2007ae}
T.~Han, Z.~Si, K.M.~Zurek and M.J.~Strassler, \emph{{Phenomenology of hidden
  valleys at hadron colliders}},
  \href{https://doi.org/10.1088/1126-6708/2008/07/008}{\emph{JHEP} {\bfseries
  07} (2008) 008} [\href{https://arxiv.org/abs/0712.2041}{{\ttfamily
  0712.2041}}].

\bibitem{Arkani-Hamed:2008kxc}
N.~Arkani-Hamed and N.~Weiner, \emph{{LHC Signals for a SuperUnified Theory of
  Dark Matter}},
  \href{https://doi.org/10.1088/1126-6708/2008/12/104}{\emph{JHEP} {\bfseries
  12} (2008) 104} [\href{https://arxiv.org/abs/0810.0714}{{\ttfamily
  0810.0714}}].

\bibitem{Baumgart:2009tn}
M.~Baumgart, C.~Cheung, J.T.~Ruderman, L.-T.~Wang and I.~Yavin,
  \emph{{Non-Abelian Dark Sectors and Their Collider Signatures}},
  \href{https://doi.org/10.1088/1126-6708/2009/04/014}{\emph{JHEP} {\bfseries
  04} (2009) 014} [\href{https://arxiv.org/abs/0901.0283}{{\ttfamily
  0901.0283}}].

\bibitem{Cheung:2009su}
C.~Cheung, J.T.~Ruderman, L.-T.~Wang and I.~Yavin, \emph{{Lepton Jets in
  (Supersymmetric) Electroweak Processes}},
  \href{https://doi.org/10.1007/JHEP04(2010)116}{\emph{JHEP} {\bfseries 04}
  (2010) 116} [\href{https://arxiv.org/abs/0909.0290}{{\ttfamily 0909.0290}}].

\bibitem{Falkowski:2010cm}
A.~Falkowski, J.T.~Ruderman, T.~Volansky and J.~Zupan, \emph{{Hidden Higgs
  Decaying to Lepton Jets}},
  \href{https://doi.org/10.1007/JHEP05(2010)077}{\emph{JHEP} {\bfseries 05}
  (2010) 077} [\href{https://arxiv.org/abs/1002.2952}{{\ttfamily 1002.2952}}].

\bibitem{Falkowski:2010gv}
A.~Falkowski, J.T.~Ruderman, T.~Volansky and J.~Zupan, \emph{{Discovering Higgs
  Decays to Lepton Jets at Hadron Colliders}},
  \href{https://doi.org/10.1103/PhysRevLett.105.241801}{\emph{Phys. Rev. Lett.}
  {\bfseries 105} (2010) 241801}
  [\href{https://arxiv.org/abs/1007.3496}{{\ttfamily 1007.3496}}].

\bibitem{CMS:2022qej}
{\scshape CMS} collaboration, \emph{{Search for long-lived particles decaying
  to a pair of muons in proton-proton collisions at $\sqrt{s}$ = 13 TeV}},
  \href{https://arxiv.org/abs/2205.08582}{{\ttfamily 2205.08582}}.

\bibitem{ATLAS:2022izj}
{\scshape ATLAS} collaboration, \emph{{Search for light long-lived neutral
  particles that decay to collimated pairs of leptons or light hadrons in $pp$
  collisions at $\sqrt{s}=13$\textasciitilde{}TeV with the ATLAS detector}},
  \href{https://arxiv.org/abs/2206.12181}{{\ttfamily 2206.12181}}.

\bibitem{CMS:2022nty}
{\scshape CMS} collaboration, \emph{{Inclusive nonresonant multilepton probes
  of new phenomena at $\sqrt s$=13\,\,TeV}},
  \href{https://doi.org/10.1103/PhysRevD.105.112007}{\emph{Phys. Rev. D}
  {\bfseries 105} (2022) 112007}
  [\href{https://arxiv.org/abs/2202.08676}{{\ttfamily 2202.08676}}].

\bibitem{ATLAS:2022vod}
{\scshape ATLAS} collaboration, \emph{{Search for Third-Generation Vectorlike
  Leptons in pp Collisions at $\sqrt {s}$ = 13 TeV with the ATLAS detector}}, .

\bibitem{Kribs:2022gri}
G.D.~Kribs, G.~Lee and A.~Martin, \emph{{Effective Field Theory of
  St\"uckelberg Vector Bosons}},
  \href{https://arxiv.org/abs/2204.01755}{{\ttfamily 2204.01755}}.

\bibitem{Aebischer:2022wnl}
J.~Aebischer, W.~Altmannshofer, E.E.~Jenkins and A.V.~Manohar, \emph{{Dark
  matter effective field theory and an application to vector dark matter}},
  \href{https://doi.org/10.1007/JHEP06(2022)086}{\emph{JHEP} {\bfseries 06}
  (2022) 086} [\href{https://arxiv.org/abs/2202.06968}{{\ttfamily
  2202.06968}}].

\end{thebibliography}\endgroup

\end{document}